\documentclass[nofootinbib,amsmath,amssymb,aps,pre,showkeys,reprint,eqsecnum]{revtex4-1}

\usepackage{graphicx}  
\usepackage{dcolumn}   
\usepackage{bm}        
\usepackage{xcolor}

\def\S{\mathcal{S}}

\def\eRM{{\mathrm e}}
\def\dRM{{\mathrm d}}

\def\mv{{\bm v}}

\def\mk{{\bm k}}

\def\eps{\varepsilon}

\allowdisplaybreaks
\allowbreak
\newcommand{\fp}[1]{FP$ {\textrm{#1}}$}

\usepackage{enumerate}
\allowdisplaybreaks

\begin{document}

\title{Turbulent compressible fluid: Renormalization group analysis, scaling regimes, and anomalous scaling of advected scalar fields}

\author{N.~V.~Antonov$^{1}$}
\email{n.antonov@spbu.ru}
\author{N.~M.~Gulitskiy$^{1}$}
\email{n.gulitskiy@spbu.ru}
\author{M.~M.~Kostenko$^{1}$}
\email{kontramot@mail.ru}
\author{T.~Lu\v{c}ivjansk\'y$^{2,3}$}
\email{tomas.lucivjansky@upjs.sk}

\affiliation{$^{1}$Department of Physics, St. Petersburg State University, 7/9 Universitetskaya nab., St. Petersburg 199034, Russia\\
$^{2}$Faculty of Sciences, P.J. \v{S}af\'arik
University, Moyzesova 16, 040 01 Ko\v{s}ice, Slovakia\\ 
$^{3}$
Peoples' Friendship University of Russia (RUDN University), 6 Miklukho-Maklaya St., Moscow, 117198, Russia}

\begin{abstract}
We study a model of fully developed turbulence of a compressible fluid, based on the stochastic Navier-Stokes equation, 
by means of the field theoretic renormalization group. 
In this approach, scaling properties are related to the fixed points of the renormalization group equations.  
Previous analysis of this model near the real-world space dimension 3 identified some scaling regime~[Theor.~Math.~Phys., {\bf 110}, 3 (1997)].
The aim of the present paper is to explore the existence of additional regimes, that could not be found using the direct perturbative approach of the previous work, and to analyze the crossover between different regimes.
It seems possible to determine them
near the special value of space dimension $4$
in the framework of double $y$ and $\varepsilon$ expansion, where 
$y$ is the exponent associated 
with the random force and $\varepsilon=4-d$  is the deviation from the space dimension $4$.
Our calculations show that there exists an additional fixed point that 
governs scaling behavior. 
Turbulent advection of a passive scalar (density) field by this velocity ensemble is considered as well.
We demonstrate that various correlation functions of the scalar field
exhibit anomalous scaling behavior in the inertial-convective range.
The corresponding anomalous exponents, identified as scaling
dimensions of certain composite fields, can
be systematically calculated as a series in $y$ and $\varepsilon$.
All calculations are performed in the leading one-loop approximation.
\end{abstract}

\pacs{05.10.Cc, 47.27.eb, 47.27.ef}

\keywords{anomalous scaling, passive scalar advection, turbulence, renormalization group}

\maketitle

\section{Introduction} \label{sec:Intro}

Understanding fully developed turbulence is a complex, rich, and challenging problem. 
Among the most important features of such  behavior are energy cascasdes
and intermittency. 
The former brings energy from large scales, responsible for the 
creation of turbulence, into smaller ones, in which viscosity plays a major role and dissipation effects dominate.
Intermittency, an irregular alternation of phases of certain dynamics, means that very rare configurations
of a system contribute most significally
to statistical distributions. In turbulence this fact manifests itself in anomalous scaling, which is characterized by 
singular behavior of various
statistical quantities as functions of the integral turbulence scales.
Anomalous scaling is thought to be related to strong fluctuations of the energy flux and, therefore, 
to deviate from the predictions of the classical Kolmogorov-Obukhov phenomenological theory~\cite{Monin,Legacy}.

Another very interesting phenomenon
is a turbulent advection of an impurity field. Both experimental studies
and numerical simulations suggest that deviations from the classical
Kolmogorov theory are even more strongly pronounced for
passively advected fields than for the velocity field itself~\cite{NonG,pass1,pass2,ShS}.
A turbulent environment may
be introduced into such models by some ``synthetic'' velocity field
with prescribed statistics or by the stochastic Navier-Stokes equation~\cite{FGV}.  Models of the former type are 
more tractable from a mathematical point of view, whereas the latter ones bear a closer resemblance to the real world.

Fully developed turbulence is characterized by existence of an inertial range~-- an interval of scales in which both the input and dissipation of energy are insignificant, and the only notable dynamical process is the re-distribution of energy along the spectrum. Therefore, one expects the inertial range to be governed by simple (and possibly universal) laws describing turbulent processes. In accordance with this  hypothesis , the classical Kolmogorov-Obukhov theory assumes that statistical characteristics of a system, i.e., its correlation and response functions, do not depend on either the internal ($l$, viscosity-related)  or the external (${\cal L}$, external force-related) scales. These assumptions lead to simple power laws for inertial range asymptotic behavior of  these functions~\cite{Monin,Legacy}.

This notwithstanding, it is well known that correlation functions can depend on the external scale due to certain kinematic effects~-- for example, the sweeping effect, in which small turbulent eddies are carried by large ones as a whole without distortion.
These kinematic effects do not influence the formation of the energy spectrum and, therefore, can be ignored in favor of Galilean invariant objects, in particular, equal-time correlation or structure functions. Nevertheless, experimental studies suggest that Galilean invariant objects also contain some dependence on ${\cal L}$, which is usually singular and is described by an infinite set of anomalous exponents~-- a phenomenon referred to as ``anomalous scaling'' and ``multiscaling.''

In many phenomenological models the anomalous exponents are related to the statistical properties of nonuniversal quantities such as the local dissipation rate, the characteristics of nontrivial structures (vortex filaments), and so on; see, e.g.,~\cite{Monin,Legacy}. Common drawbacks of such models are that they are only loosely based on the underlying hydrodynamical equations, and they involve arbitrary adjustment parameters; therefore, these models cannot be used to construct systematic perturbation theory in a small expansion parameter~\cite{DeDom2}.

Therefore, an essential goal of the theory of turbulence is 
to construct an analytic framework based on a dynamical model, e.g., the Navier-Stokes equation.
The key obstacle in this situation
is the lack of a small expansion parameter~-- at least, a formal one. In ordinary perturbation theory for the stochastic Navier-Stokes equation (i.e., expansion in the nonlinearity) the actual expansion parameter reduces to the Reynolds number, which tends to infinity for the developed turbulence.

A similar problem is longstanding 
in the theory of critical phenomena, where it was successfully solved by use of the renormalization group (RG) method borrowed from quantum field theory. The RG method performs a certain rearrangement (infinite resummation) of the original perturbation series, and turns them into a series of the parameter of order unity. Typically, that parameter is $\eps=4-d$, the deviation of the dimensionality of space $d$ from its upper critical value~\cite{Zinn,Vasiliev,turbo,Amit,Tauber,Shirkovl}, hence the term ``epsilon expansion.'' Such expansions are still divergent, but they allow one to prove the existence of infrared (IR) scaling behavior (if such exists)
and to systematically calculate the corresponding dimensions as series in $\eps$.

The RG method and $\eps$ expansion are equally applicable to the stochastic Navier-Stokes equation if the correlation function of the random force is chosen as a power function of  the momentum $k=|\mk|$ in the form of $k^{4-d-y}$, see~\cite{turbo,HHL} and  Sec.~\ref{sec:Model} in the present paper. Here, $d$ becomes a free parameter, while the role of the RG expansion parameter is played by the exponent $y$.

The results, obtained within the $\eps$ expansions, are reliable for asymptotically small values of $\eps$, but it is unclear whether these results can be extrapolated to finite (and non-small) realistic values of $\eps$.
For this reason, in the theory of critical phenomena, 
one tries to calculate as many terms of the $\eps$ expansion as possible, to get their higher-order asymptotic form (using the instanton calculus) and to use additional methods of summation (for instance, Pad\'e-Borel or Leroy-Borel transformations). 
A common opinion is that the $\eps$ expansion indeed works for  real $\eps$~\cite{Zinn,Vasiliev,Amit,Shirkov}.

For the Navier-Stokes turbulence, the situation is much more difficult. First, the RG expansion parameter $y$ is not small, and the RG series are divergent. Second, the higher-order calculations are extremely cumbersome. Third, the higher-order asymptotic forms of the coefficients are not known yet.
However, these problems can be considered as being of technical (calculational) nature. There are more serious problems, specific only for turbulence, which are related to real physical effects: sweeping of small turbulent eddies by large-scale ones, and the anomalous scaling. In perturbation theory both of them manifest as strong divergences of the perturbation diagrams at ${\cal L}\to\infty$ (where ${\cal L}$ is the integral, i.e., external, turbulence scale) in the inertial range. Adequate analysis of these issues takes one far beyond the standard RG method: the method should be combined with the short-distance operator product expansion (OPE). 

The feature specific of the models of turbulence is the existence in the corresponding OPE of composite fields (``operators'' in quantum field terminology) with negative critical dimensions. These operators (termed ``dangerous'') give rise to strong IR singularities in the correlation functions; see~\cite{JETP,LOMI,JphysA,turbo, Vasiliev}.
While experimental data suggests that in the inertial range correlation and structure functions exhibit anomalous scaling, it has not been possible to demonstrate this property through theoretical modeling.
The main problem is the following:  if a dangerous operator is present in some field theory
there are, in fact, infinitely many such operators; moreover, the spectrum of these operators is not bound from below. Thus, there is no main, or ``most dangerous,'' operator in the model that would provide the main contribution in the corresponding OPE; see, e.g., Appendix~A in~\cite{AK15}. Therefore, the problem requires one to perform the explicit construction of all invariant scalar operators with negative dimensions, the exact calculation of their critical dimensions, and the (infinite) summation of their contributions in the corresponding OPE.

Clearly, there is little hope to solve this problem in the forseeable future.
Fortunately, situation simplifies for two important cases: sweeping effects and passive advection.
The first example is provided by the composite operators, which are powers of the velocity field in the stochastic Navier--Stokes equation. Owing to the Galilean symmetry of the model, their dimensions can be found exactly, and their contributions into the OPE can be summed up into an explicit closed expression~\cite{JETP,LOMI,turbo, Vasiliev}. This gives adequate description of the sweeping effects within the RG+OPE approach.

The second case is provided by the passive advection of scalar or vector fields by a given velocity statistics. 
The stochastic advection-diffusion equation is linear in these fields, therefore, only finite number of dangerous operators contribute to the OPE 
for any given correlation function, and the additional resummation of the series, discussed above, is not required~\cite{RG}. Therefore, one way to investigate 
these phenomena is to consider a passive advection of different types of fields by succeedingly more complex and realistic turbulent velocity environments.
In  a number of papers the RG+OPE approach has been applied to passive advection by 
Kraichnan's ensemble (the velocity field is assumed to be isotropic, Gaussian, not correlated in time,
and to have a power-like correlation function; the fluid is assumed to be incompressible)~\cite{Kraich1,GK,RG,HHL},
and by the ensemble's numerous generalizations: large-scale anisotropy, helicity,
compressibility, finite correlation time, non-Gaussianity, and a
more general form of nonlinearity~\cite{cube,A99,V96-mod,amodel,TWP,AG12,J13,Uni,Uni2,Uni3,VectorN,JJ15}.
This approach can be generalized to a non-Gaussian velocity field governed by the stochastic Navier-Stokes equation, 
to study both the velocity field's scaling behavior and passive impurity fields it advects~\cite{NSpass,Ant04,AK14,AK15,AGM,JJR16}.
The main advantage of the RG+OPE approach as applied to turbulence is that it is 
based on a microscopic model and, therefore, allows one to construct a systematic perturbation expansion for the anomalous exponents.

Until now, the majority of studies on fully developed turbulence have been concerned with incompressible fluid. 
Nevertheless, several results for the problems of universality and scaling for compressible fluids have also been obtained~\cite{El,VM,Avell,tracer2,tracer3,tracer,RG1,AH,AG,AKens,AK2}. All of them 
hint at large influence  of compressibility both on the velocity field itself and on passively advected quantities. 
In particular, a transition  from
a turbulent to a certain purely chaotic state may occur at 
large degrees of compressibility~\cite{tracer}.
In other papers corrections in the Mach number to the incompressible scaling regime were studied~\cite{LM,KompOp,NV}. 
The main result of those studies is that obtained corrections become arbitrarily large and destroy the incompressible scaling regime for 
fixed Mach numbers and large distances, what can be explained by existence of a crossover to another, yet unknown regime.
The studies~\cite{ANU97,St,MTY} were devoted to compressible fluids. The results are rather controversial; particularly, 
the model, considered in~\cite{St}, appears to be non-renormalizable.
From a general point of view, further investigations of compressibility are therefore called for.

In this paper we present an application of the field
theoretic renormalization group to the scaling regimes of a compressible fluid whose
behavior is governed by a proper generalization of the stochastic Navier-Stokes equation~\cite{ANU97}.
The stationary scaling regimes in this approach
are associated with IR attractive fixed points of the
corresponding multiplicatively renormalizable field theoretic models.  
One nontrivial fixed point for this model, attractive in our region of interest, was found in~\cite{ANU97}.
In~\cite{AK14,AK15} the scaling properties of passively advected scalar and vector fields by this velocity ensemble were investigated and anomalous scaling for different correlation functions was discovered.

The analysis in~\cite{ANU97} and subsequent papers~\cite{AK14,AK15} is self-contained and 
internally consistent for asymptotically small $y$, where 
$y$ describes the scaling behavior of a random force [see~\eqref{power}]. 
However, as $y$ grows some effects can happen: the fixed point, found within the analysis at small $y$, can lose its stability or go to unphysical region. In the same time, another fixed point(s), which are not  ``visible'' within the $y$-expansion, may begin to determine the IR behaviour.
Therefore, it is feasible that other fixed points (and other scaling regimes with other critical dimensions) exist for finite (non-small) values of $y$.  These fixed points cannot be identified within the frameworks of the analysis at small $y$ and in this sense are non-perturbative. However,  some of them can be revealed in a double expansion in $y$ and the deviation  from the space dimension $d$ from some exceptional values, like $d=2$ for the incompressible case~\cite{HN96,Two,AHKV05}. Indeed, for the incompressible case, the double expansion in $y$ and $(d-2)$ reveals two non-trivial fixed points (and hence two asymptotic regimes), one corresponding to the equilibrium (thermal) regime~\cite{FNS} and the other to the turbulence~\cite{HN96,turbo}.

In the compressible case there are also two special dimensions, namely $d=2$ and $d=4$, in which the renormalization procedure is much more complicated in comparison to all other situations.
The double expansion around $d=2$ is currently under consideration~\cite{Tomas}; 
$d=4$ admits the double expansion in $y$ and $\varepsilon=(4-d)$ and is
employed in the present paper. 
Model analysis near this special dimension allows us to not only extend previous research~\cite{ANU97,AK14} (by refining the known scaling regimes through resummation of the ordinary $y$ expansion), but also to investigate the existence of other possible regimes.
This is the main subject and motivation of the present  study.  We show that a new fixed point (henceforth called ``local'' for the reasons to be explained) indeed exists near $d=4$ and persists, at least in the leading one-loop approximation, for all $d$. 
The other ``non-local'' fixed point corresponds to the scaling regime found earlier~\cite{ANU97}. The regions of stability and critical dimensions 
for both fixed points are calculated in the leading one-loop order.

Following the procedure of previous {studies~\cite{ANU97,AK14,AGKL1}}, first the stochastic Navier-Stokes equation is discussed. 
After establishing the existence of necessary fixed points of Navier-Stokes equation, the advection of 
scalar fields is explored.
Since RG functions of the parameters, entering the Navier-Stokes equation do not depend on the parameters, connected
with advection-diffusion equation, this is a possible and probably the easiest approach.

The paper is organized as follows:
\newline

In Sec.~\ref{sec:Model}, a detailed description of the stochastic Navier-Stokes equation for a compressible fluid is given.
Sec.~\ref{sec:QFT} is devoted to field theoretic formulation of the model
and the corresponding diagrammatic technique. In particular, possible types of divergent Green functions at $d=3$ and $d=4$ and the
necessity of introducing a new coupling constant at $d=4$ are discussed.
In Sec.~\ref{sec:Renorm}, the renormalizability of the model is established and one-loop explicit
expressions for the renormalization constants and RG functions (anomalous dimensions and $\beta$ functions) are derived. 
In Sec.~\ref{sec:RC}, the obtained expressions for RG functions are examined. 
IR asymptotic  behavior, obtained by solving the RG equations, is discussed. It is shown that, depending  on two exponents $y$ and $\varepsilon$, the 
RG equations possess an 
IR attractive fixed point, which implies existence of a scaling regime in the inertial range. The corresponding scaling dimensions of all fields
and parameters of the model are presented.

In Sec.~\ref{sec:Ops1}, an advection of a passive scalar (density) field by compressible velocity field which obeys Navier-Stokes equation is analyzed. 
A field theoretic formulation of  the full model is presented. 
It is shown that the full model is multiplicatively renormalizable; 
the existence of a scaling regime in the IR range is established. 
The renormalization of composite operators is carried out. An inertial-range behavior of various correlation functions is studied
by means of the OPE. 
It is shown that leading terms of the inertial-range behavior are determined by the contributions of the operators built solely from the scalar
fields. 
As a result, the IR behavior of the pair correlation functions of the composite operators is power-like with negative critical 
dimensions~-- a situation, called anomalous scaling.

Sec.~\ref{sec:Conc} is reserved for conclusions.
The main one is that the new (local) fixed point indeed exists in the model of turbulence for a compressible fluid, based
on the stochastic Navier-Stokes equation.

Appendices~\ref{app1} and~\ref{app2} contain detailed calculations of all diagrams, needed to perform multiplicative renormalization of our model.

\section{Description of the model} \label{sec:Model}

The Navier-Stokes equation for a viscous compressible fluid can be written in the following form~\cite{LL}:
\begin{eqnarray}
\rho\nabla_{t} v_{i} = \nu_{0} (\delta_{ik}\partial^{2}-
\partial_{i}\partial_{k}) v_{k}
+ \mu_0 \partial_{i}\partial_{k} v_{k} - \partial_{i} p + \eta_{i},
\nonumber \\ {}
\label{NS}
\end{eqnarray}
where the differential operator in the right hand side
\begin{equation}
\nabla_{t} = \partial_{t} + v_{k} \partial_{k}
\label{nabla}
\end{equation}
is the Lagrangian (convective) derivative, 
$\rho$ is a fluid density  field, $v_i$ is the velocity field,
$\partial_{t} =
\partial /\partial t$, $\partial_{i} = \partial /\partial x_{i}$, 
$\partial^{2} =\partial_{i}\partial_{i}$ is the Laplace operator, $p$ is the pressure field,  
and $\eta_i$ is the density of an external force per unit volume. 
The fields $v_i$, $ \eta_i$, $\rho$ and $p$ depend on  $x=(t,{\bm x})$ with
${\bm x}=(x_1,x_2,\ldots,x_d)$, where
$d$ is the dimensionality of space.
The constants $\nu_{0}$ and $\mu_{0}$ are two
independent molecular viscosity coefficients~\cite{LL}; in~\eqref{NS} we 
have explicitly separated the transverse and longitudinal
components of the viscous term. Summations over repeated vector indices are always implied in this work.

To get a closed system of equations, the model~(\ref{NS}) 
must be augmented by two additional equations, namely a continuity equation and an equation of state between 
deviations $\delta p$ and $\delta \rho$ from the equilibrium values. They explicitly read
\begin{eqnarray}
\partial_{t} \rho  + \partial_{i} (\rho v_{i})   = 0
\label{CE}
\end{eqnarray}
and
\begin{eqnarray}
  \delta p = c_0^2 \delta\rho.
\label{SE}
\end{eqnarray}

In order to derive renormalizable field theory, the  stochastic equation~(\ref{NS}) has to be 
divided by $\rho$, and fluctuations in viscous terms have to be neglected~\cite{NV}. Further, by using the expressions~\eqref{CE} and~\eqref{SE}
the problem can be recast in the form of two coupled equations:
\begin{eqnarray}
\nabla_{t} v_{i} &=&
\nu_{0} (\delta_{ik}\partial^{2}-\partial_{i}\partial_{k})
v_{k}\! +\! \mu_0 \partial_{i}\partial_{k} v_{k} -\!
\partial_{i} \phi\! +\! f_{i},
\label{ANU} \\
\nabla_{t} \phi &=& -c_{0}^{2} \partial_{i}v_{i}.
\label{ANU1}
\end{eqnarray}
Here, a new scalar field $\phi=\phi(x)$ is related to the density fluctuations via the relation $\phi = c_0^2 \ln (\rho/\overline{\rho})$.
A parameter $c_0$ is the adiabatic speed of sound, $\overline{\rho}$ denotes the mean value of $\rho$, and 
$f_{i}=f_{i}(x)$ is a density of the external force per unit mass.

In the stochastic formulation of the problem the turbulence is modeled by an external force~-- {it} is assumed to be a random variable, which
mimics the input of energy into the system from the outer large scale ${\cal L}$. Its precise
form is believed to be unimportant and is usually considered to be a random Gaussian variable with zero mean
and prescribed correlation function~\cite{Vasiliev}. For the use of the
standard RG technique this correlator must exhibit
a power law asymptotic behavior at large wave numbers~\cite{turbo,China}.
In the case of compressible fluid it should be naturally augmented with a longitudinal component, hence, the simplest way is to choose
it in the form~\cite{ANU97}
\begin{equation}
  \langle f_i(t,{\bm x}) f_j(t',{\bm x}') \rangle= \frac{\delta(t-t')}{(2\pi)^d} \int_{k>m} \dRM^d{\bm k}\mbox{ } \widetilde{D}_{ij}({\bm k})
  \eRM^{i{\bm k}\cdot({\bm x-\bm x'})},
\label{ff}
\end{equation}
where the argument is given by
\begin{equation}
  \widetilde{D}_{ij}({\bm k}) = g_{10} \nu_0^3 k^{4-d-y} \biggl\{  P_{ij}({\bm k}) + \alpha Q_{ij}({\bm k}) \biggl\}.
  \label{power}
\end{equation} 
Here, $P_{ij}({\bm k}) = \delta_{ij}-k_ik_j/k^2$ and $Q_{ij}({\bm k})=k_ik_j/k^2$ are the transverse 
and longitudinal
projectors, $k=|{\bm k}|$, the amplitude $\alpha$ is a free parameter,
the amplitude $g_{10}$ is a coupling 
constant (formal expansion parameter in the ordinary perturbation theory);
the relation $g_{10} \sim \Lambda^{y}$ sets in the typical ultraviolet (UV) momentum scale $\Lambda$, which is
a reciprocal of the dissipation length scale.
A parameter $m={\cal L}^{-1}$ provides an infrared 
regularization; its precise form is unessential and the sharp
cut-off is the simplest choice for calculation purposes. 
The exponent $y$ provides analytic UV regularization and,
therefore, plays a role of a formally small expansion parameter~\cite{Vasiliev}.
The most realistic (physical) value is obtained in the limit
$y\to4$, when the function in~(\ref{power}) can be interpreted as power-like representation of the
Dirac function $\delta({\bm k})$: physically it corresponds to the idealized picture of the energy
input from infinitely large scales. The Galilean 
invariance for the model~(\ref{NS}) is ensured when the function~\eqref{power} is delta-correlated in time~\cite{turbo}.

\begin{widetext}
\section{Field theoretic formulation of the model}\label{sec:QFT}

\subsection{Action functional and Feynman rules} \label{sec:Diag}

According to the general theorem~\cite{Vasiliev,Zinn}, the stochastic problem~\eqref{ANU}~-- \eqref{power} is equivalent to the field theoretic
model with a doubled set of fields $\Phi=\left\{v_{i}, v_{i}',\phi, \phi'\right\}$ and De Dominicis-Janssen action functional, written
in a compact form as
\begin{align}
  \S_{\bm v}(\Phi) & = \frac{v_i' \widetilde{D}_{ij} v_j'}{2} 
  +v_i' \biggl[
  -\nabla_t v_i + \nu_0(\delta_{ij}\partial^2 - \partial_i \partial_j)v_j
  +u_0 \nu_0 \partial_i \partial_j v_j - \partial_i \phi
  \biggl] \nonumber \\  &
  +\phi'[-\nabla_t \phi + v_0 \nu_0 \partial^2 \phi - c_0^2 (\partial_i v_i)],
  \label{eq:action} 
\end{align}
\end{widetext}
where $\widetilde{D}_{ij}$ is the correlation function~(\ref{power}).
Here we employ a condensed notation, in which integrals over the spatial variable 
${\bm x}$ and the time variable $t$, as well as summation over repeated indices, are implicitly assumed, i.e.,
\vspace{-.5em}
\begin{eqnarray}
\label{quadlocal2}
{\phi'}\partial_t{\phi}&=&\int \dRM t\int \dRM^d{\bm x}\, \phi'(t,{\bm x})\partial_t\phi(t,{\bm x}); \\ \nonumber 
v'_iD_{ik}{v'}_k&=&\int \dRM t \!\int\! \dRM^d{\bm x} \!\! \int\! \dRM^d{\bm x'} v_i(t,{\bm x})D_{ik}({\bm x}-{\bm x'})v_k(t,{\bm x'}). 
\end{eqnarray}

Moreover, we have introduced a new dimensionless parameter
$u_0=\mu_0/\nu_0>0$ and a new term
$v_0 \nu_{0} \phi' \partial^{2}\phi$ with another positive
dimensionless parameter $v_0$, which is needed  to ensure 
multiplicative renormalizability. 
The action~(\ref{eq:action}) is amenable to the standard methods
of the quantum field theory, such as Feynman diagrammatic technique and 
renormalization group procedure.

In the standard field theoretic approach to the stochastic Navier-Stokes equation, the actual RG expansion parameter is $y$, while $d$ plays a passive role;
see the monographs~\cite{turbo,Vasiliev} for details. 
Our approach closely follows the analysis of the 
incompressible Navier-Stokes equation near
space dimension $d=2$~\cite{HN96,Two,AHKV05}. In this case
an additional UV divergence appears in the Green function $\left\langle v'_iv'_j\right\rangle$. It can
be absorbed by a suitable local counterterm $v'_i\partial^2 v'_i$, and a regular
expansion in both $y$ and $\eps'=d-2$ must be constructed. 
Up to now the present model~(\ref{eq:action}) has been 
investigated in fixed space dimension $d=3$, for which the action~(\ref{eq:action}) contains all terms that can be 
generated during the renormalization procedure~\cite{ANU97,AK14,AK15}. 
However, from the dimensional analysis (see below) it follows that in $d=4$ an additional divergence appears in a similar fashion
in the Green function $\left\langle v_i'v_j'\right\rangle$. Therefore, to keep the model renormalizable in $d=4$, the kernel function $\widetilde{D}_{ij}({\bm k})$ 
in~(\ref{ff}) has to be {replaced by} $D_{ij}({\bm k})$, where
\begin{equation}  
  D_{ij}({\bm k})=g_{10} \nu_0^3 k^{4-d-y} \biggl\{
  P_{ij}({\bm k}) + \alpha Q_{ij}({\bm k})
  \biggl\} 
  + g_{20} \nu_0^3 \delta_{ij}.
  \label{eq:correl2}
\end{equation}
A new term on the right hand side with an additional coupling constant $g_{20}$  absorbs divergent contributions from $\left\langle v_i'v_j'\right\rangle$. 
In contrast to the two-dimensional incompressible case~\cite{HN96} no momentum dependence is needed here.

The field theoretic formulation~(\ref{eq:action}) means that various correlation 
and response functions of the original stochastic problem are
represented by functional averages over the full set of fields with the functional weight
$\exp {\cal S}(\Phi)$, and in this sense they can be viewed as Green
functions of the field theoretic model \cite{Vasiliev,Zinn}.

\begin{figure}[t]
   \centering
   \begin{tabular}{c}
     \includegraphics[width=6.1cm]{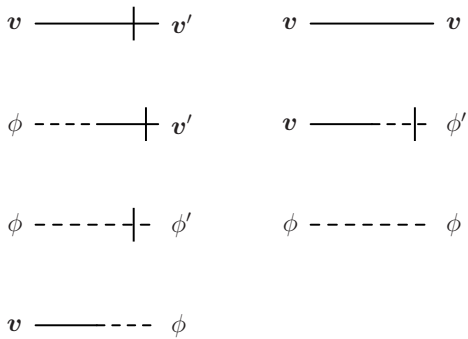}
      \end{tabular}
   \caption{Graphical representation of the bare propagators in the model~(\ref{eq:action}).}
   \label{fig:prop_vertex}
\end{figure}

\begin{figure}[t]
   \centering
   \begin{tabular}{c c}
     \includegraphics[width=6.1cm]{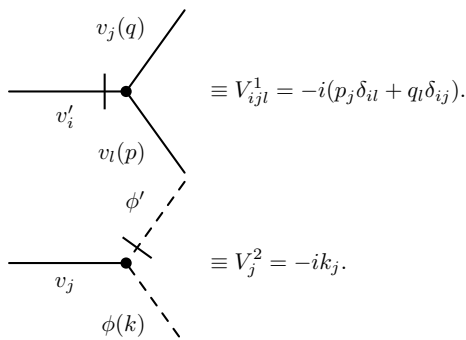}  
      \end{tabular}
   \caption{Graphical representation of the interaction vertices in the model~(\ref{eq:action}).}
   \label{fig:2}
\end{figure}

The perturbation theory of the model can be {constructed} according to the usual 
Feynman diagrammatic expansion~\cite{Zinn,Vasiliev}. 
Bare propagators are read off from the inverse matrix of the Gaussian (free) part of the action functional, while
a nonlinear part of the differential equation {determines} the interaction vertices.
A graphical representation of the propagator functions is depicted in Fig.~\ref{fig:prop_vertex}, and of
the vertices~-- in Fig.~\ref{fig:2}.
From~\eqref{eq:action} it follows that the Feynman diagrammatic technique for this model contains two
interactions, $-v_i'(v_j\partial_j)v_i$ and $-\phi'(v_i\partial_i)\phi$. The propagator functions in 
the frequency-momentum representation read
\begin{eqnarray}
\langle v_iv_j' \rangle_{0} &=& \overline{\langle v_j'v_i \rangle}_{0}= P_{ij}({\bm k})
\epsilon_{1}^{-1} + Q_{ij}({\bm k}) \epsilon_{3} R^{-1} ,
\nonumber \\
\langle v_iv_j \rangle_{0} &=&   {P_{ij}({\bm k})} \frac{d_1^{f}}{|\epsilon_{1}|^{2}}
+  {Q_{ij}({\bm k})} d_2^{f} \left|\frac{\epsilon_{3}}{R}\right|^{2},
\nonumber \\
 {\langle \phi v_j' \rangle_{0}} &=&  {\overline{\langle v_j' \phi \rangle}_{0}}= -
\frac{{\rm i}c_{0}^{2} {k_j}}{R}, \quad
 {\langle  v_i\phi' \rangle_{0}} =  {\overline{\langle \phi'v_i \rangle}_{0}}= -\frac{{\rm i} {k_i}}{R},
\nonumber \\
\langle  \phi\phi' \rangle_{0} &=&  {\overline{\langle \phi'\phi \rangle}_{0}}=
\frac{\epsilon_{2}}{R}, \hskip1.25cm
\langle  \phi\phi \rangle_{0} = \frac {c_{0}^{4} k^{2}d_2^{f}}
{|R|^{2}},
\nonumber \\
 {\langle  v_i\phi \rangle_{0}} &=&  {\overline{\langle  \phi v_i \rangle}_{0}} =
\frac{ {\rm i} c_{0}^{2} d_2^{f}\epsilon_{3}  {k_i}} {|R|^{2}} ,
\nonumber \\
\langle  \phi'\phi' \rangle_{0} &=&  {\langle  v_i'\phi' \rangle_{0}} =
 {\langle  v_i'v_j' \rangle_{0}} = 0,
\label{lines}
\end{eqnarray}
 {where the symbol $\overline{z}$ denotes the complex conjugate of the expression
$z$.}
For convenience, the following abbreviations have been used
\begin{align}
\epsilon_{1}  & = -{\rm i}\omega +\nu_0 k^{2}, 
&\epsilon_{2}&=-{\rm i}\omega+ u_0\nu_0 k^{2} ,
\nonumber \\
\epsilon_{3}  & =  -{\rm i}\omega+ v_0\nu_0 k^{2} ,  
&R&=\epsilon_{2}\epsilon_{3}+c_{0}^{2}k^{2}
\end{align}
and
\begin{align}
d^{f}_1  &=  g_{10}\nu_0^{3}\, k^{4-d-y}+g_{20}\nu_0^3, \nonumber \\
d^{f}_2  &=  \alpha  g_{10}\nu_0^{3}\, k^{4-d-y}+g_{20}\nu_0^3.
\label{energies}
\end{align}

In the limit $c_{0}\to\infty$ the propagator functions  {$\langle  v_iv_j \rangle_{0}$} and  {$\langle  v_iv_j' \rangle_{0}$} become purely 
transverse,  and all the mixed propagators except  {$\langle  \phi v_j' \rangle_{0}$} vanish.  Moreover, the scalar fields $\phi$ and $\phi'$ decouple 
from the velocity fields  {$v_i$} and  {$v'_i$}~-- it is impossible to construct a diagram with 
only velocity fields $ {v_i}$ and $ {v_i'}$ as external lines, containing internal lines with fields $\phi$ or $\phi'$.
Thus, in conformity with the physical point of view, the well-known Feynman rules for the incompressible
fluid~\cite{Vasiliev,turbo} are obtained.

 {From here on, a solid line without a slash denotes the field $v_i$, a solid line with a slash corresponds to the field $v_i'$,
a dashed line without a slash denotes the field $\phi$, and a dashed line with a slash corresponds to the field $\phi'$.}

\subsection{Canonical dimensions, UV divergences, and renormalization constants} 
\label{sec:Canon}

Ultraviolet renormalizability is very efficiently exhibited in analysis
of the 1-particle irreducible Green functions, later referred to as 1-irreducible Green functions following the notation in~\cite{Vasiliev}. 
In the case of dynamical models \cite{Vasiliev,Tauber}
two independent scales have to be introduced: the time scale $T$ and the length scale $L$. Thus
the canonical dimension of any quantity $F$ (a field or a parameter) is
described by two numbers, the
frequency dimension $d_{F}^{\omega}$ and the momentum dimension $d_{F}^{k}$,
defined such that
\begin{align}
  d_k^k&=-d_{ x}^k=1, &d_k^{\omega}& =d_{x}^{\omega }=0,\nonumber\\
  d_{\omega }^{\omega }&=-d_t^{\omega }=1, &d_{\omega }^k&=d_t^k=0,
  \label{def_dim}
\end{align}
i.e.,
\begin{eqnarray}
[F] \sim [T]^{-d_{F}^{\omega}} [L]^{-d_{F}^{k}}.
\label{canon}
\end{eqnarray}
The remaining dimensions can be found from the requirement that each term of the
action functional be dimensionless, with respect to the momentum and the
frequency dimensions separately.

Based on $d_F^k$ and $d_F^\omega$  the total canonical dimension
$d_F=d_F^k+2d_F^\omega$ can be introduced, which in
the renormalization theory of dynamic models plays the
same role as the conventional (momentum) dimension does in
static problems. 
 {Setting $\omega \sim k^{2}$ ensures that all the viscosity and diffusion coefficients in the model are dimensionless. }
Another option is to set the speed of sound $c_{0}$
dimensionless and  consequently obtain that $\omega \sim k$, i.e., $d_{F}=d_{F}^{k}+d_{F}^{\omega}$.
This variant would mean that
we are interested in the asymptotic behavior of the Green functions
as $\omega \sim k \to 0$, in other words, in sound modes in turbulent medium. Even though this problem
is very interesting itself, it is not yet accessible {for} the
RG treatment, so we will not discuss it here.
The choice $\omega \sim k^{2} \to 0$ is the same as in the models 
of incompressible fluid, where
it is the only possibility because the speed of sound is infinite.
A similar alternative in dispersion laws exists, for example, within the so-called model H of equilibrium
dynamical critical behavior, see~\cite{Vasiliev,Tauber}.

The canonical dimensions for the model~(\ref{eq:action}) are listed in
Table~\ref{table1}, including renormalized parameters (without the subscript ``0'') and scalar impurity fields $\theta$ and $\theta'$ {and parameter $w$}, 
which appears in Sec.~\ref{sec:Ops1}. From Table~\ref{table1} it follows that the model
is logarithmic (the coupling constants $g_{10} \sim [L]^{-y}$
and $g_{20} \sim [L]^{-\varepsilon}$
become dimensionless) at $y=\varepsilon=0$.
In this work we use the minimal subtraction (MS) scheme for the calculation
of renormalization constants. In this scheme the UV divergences in the
Green functions manifest
themselves as poles in $y$, $\varepsilon$ and their linear combinations. Here, in accordance
with critical phenomena we retain the notation $\varepsilon=4-d$ .

\begin{table*}
\caption{Canonical dimensions of the fields and parameters.}
\label{table1}
\begin{ruledtabular}
\begin{tabular}{c|c|c|c|c|c|c|c|c|c|c|c|c}
$F$ & $ {v_i'}$ & $ {v_i}$ & $\phi'$ & $\phi$ & $ \theta' $ &  $\theta$ &
$m$, $\mu$, $\Lambda$ & $\nu_0$, $\nu$, $\kappa$, $\kappa_0$  & $c_{0}$, $c$ &
$g_{10}$ & $g_{20}$ & $u_{0}$, $v_{0}$ $w_{0}$, $u$, $v$, $w$, $g_1$, $g_2$, $\alpha$  \\
\hline
$d_{F}^{k}$ & $d+1$ & $-1$ & $d+2$ & $-2$ & $d$  & 0
& 1 &  $-2$ & $-1$ & $y$ & $4-d$ & 0 \\
$d_{F}^{\omega}$ & $-1$ & 1 & $-2$ & 2 & $1/2$ & $-1/2$ &
0 & 1 & 1 & 0 & 0 & 0\\
$d_{F}$ & $d-1$ & 1 & $d-2$ & 2 & $d+1$ & $-1$ & 1 & 0 & 1 & $y$ & $4-d$ & 0 \\
\end{tabular}
\end{ruledtabular}
\end{table*}

The total canonical dimension of any 1-irreducible Green function $\Gamma$ is expressed by the relation
\begin{eqnarray}
\delta_{\Gamma} = d+2 - \sum_{\Phi} N_{\Phi} d_{\Phi},
\label{index}
\end{eqnarray}
where $N_{\Phi}$ is the number of the given type of field {entering the function}
$\Gamma$, $d_{\Phi}$ is the corresponding total canonical dimension of field $\Phi$, and
the summation runs
over all types of the fields $\Phi$ in function $\Gamma$~\cite{Vasiliev,Zinn,Tauber}.  

Superficial UV divergences whose removal requires counterterms can be present only in
those functions $\Gamma$ 
for which
the formal index of divergence $\delta_{\Gamma}$ is a non-negative integer.
Dimensional analysis should be augmented by the following additional considerations:
\begin{enumerate}[(1)]
\item In any dynamical model of the type~\eqref{eq:action} all the {1-irreducible} functions without the response fields $ {v_i'}$ or $\phi'$
necessarily contain closed circuits of retarded propagators. Therefore, such functions
vanish identically, i.e., they do not require counterterms.
\item The field $\phi$ enters the vertex  {$\phi'(v_i\partial_i)\phi$}
only in the form of  a spatial derivative, 
which reduces the real index of
divergence:
\begin{equation}
\delta_{\Gamma}' = \delta_{\Gamma}- N_{\phi}.
\label{real}
\end{equation}
In particular, this means that the field $\phi$ enters
the counterterms only in the form of the derivative  {$\partial_i\phi$}. In fact, not all 
counterterms allowed by dimensional analysis are present. For example, for
the 1-irreducible function
$\langle\phi'\phi\rangle$
one obtains $\delta_{\Gamma} = 2$, $\delta_{\Gamma}' = 1$, thus, the
only possible counterterm is
$\phi' \partial^{2}\phi$, while the 
structure $\phi' \partial_{t}\phi$
is forbidden.
\item Since the random noise~\eqref{ff} is white in-time, the model~(\ref{eq:action}) is Galilean invariant. Hence, the
contributions of the counterterms have to respect this invariance. In particular, 
the covariant derivative~(\ref{nabla}) must enter the
counterterms as a whole. This  imposes some restrictions on 
possible counterterms: the counterterm required
for the {1-irreducible} function  {$\langle \phi' v_i \phi\rangle$} with
$\delta_{\Gamma} = 1$, $\delta_{\Gamma}' = 0$, necessarily attains the form
 {$\phi'(v_i\partial_i)\phi$} and can appear only in the combination
$\phi'\nabla_{t}\phi$ with the counterterm $\phi' \partial_{t}\phi$
discussed above. Hence, it is forbidden.

\item An additional observation which reduces possible types of counterterms is the 
generalized Galilean invariance with the time-dependent vector
transformation velocity parameter ${\bm w}(t)$:
\begin{align}
  {\bm v}_{w}(x) & =  {\bm v}(x_{w}) - {\bm w}(t), 
  &x &= (t,{\bm x}),
  \nonumber \\
    {\Psi_{w}(x)}& =  {\Psi(x_{w})};  &x_{w}& =(t,{\bm x}+{\bm u}(t));
  \nonumber \\
  {\bm u}(t) & = \int^{t}_{ -\infty} {\bm w}(t') \, \dRM t',
  \label{GGi}
\end{align}
where  {$\Psi$} stands for any of the three remaining fields~-- $ {v_i'},\phi',\phi$.
The crucial idea is that {despite} the fact that the action functional
is not invariant  with respect to such a transformation, it transforms in the identical way as 
the generating functional
of the {1-irreducible} Green functions:
\begin{eqnarray}
  {\cal S}_\mv ( {\Psi_{w}}) &=&
   {\cal S}_\mv( {\Psi}) +  {v_i'\partial_{t} w_i}, \nonumber \\
{\Gamma}( {\Psi_{w}}) &=& {\Gamma}( {\Psi}) +  {v_i'\partial_{t} w_i}.
\label{gGal}
\end{eqnarray}
Since the latter formula can be rewritten in the form
\begin{equation}
  \Gamma( {\Phi}) =  \S ( {\Phi}) + \widetilde{\Gamma}( {\Phi}),
  \label{eq:expansion1}
\end{equation}
where  {$\Phi$} is the set of all the fields, $ {\Phi}=\left\{ { v_i}, { v_i'},\phi,\phi' \right\}$,
${\S}( {\Phi})$ is the given action functional, and $\widetilde{\Gamma}( {\Phi})$ is the sum of all the {1-irreducible} loop diagrams 
that contain all the UV divergences.  The expressions~\eqref{gGal} mean that the counterterms appear invariant under
the generalized Galilean transformation~(\ref{GGi}).

The above considerations exclude the counterterm  {$v_i'\nabla_{t}v_i$}, invariant
with respect to the conventional Galilean transformation with a constant vector
${\bm w}$, but not invariant with respect to~(\ref{GGi}). In particular, 
the only possible counterterm for {1-irreducible} function  {$\langle v_i'v_j\rangle$}
with $\delta_{\Gamma} = 2$ is  {$v_i'\partial^{2} v_i$},
and that the 1-irreducible function  {$\langle v_i'v_jv_k\rangle$} with $\delta_{\Gamma} = 1$ does not diverge. 
More detailed discussions of the application of the generalized Galilean transformation can be found in~\cite{DeDom,DeDom2,turbo,Vasiliev,GGp}.

\item From the expressions~(\ref{lines}) for propagators it follows that 
propagators containing the field $\phi$, namely,
 {$\langle v_i'\phi \rangle_{0}$},  {$\langle v_i \phi \rangle_{0}$}, and $\langle \phi\phi \rangle_{0}$, 
contain the factors $c_{0}^{2}$ or $c_{0}^{4}$. Since $d^k_c\neq0$ and $d^\omega_c\neq0$,  parameter $c_0$ shows up as an external numerical factor in any diagram
involving these propagators, and its real index of divergence reduces by
the corresponding number of unities. 
In particular, any diagram of the 1-irreducible function with $N_{\phi'}>N_{\phi}$ must contain the factor
$c_{0}^{2(N_{\phi'}-N_{\phi})}$. It then follows that the counterterm to
the {1-irreducible} function  {$\langle \phi'v_j \rangle$} with
$\delta_{\Gamma}=3$ inevitably reduces to  {$c_{0}^{2}\phi' (\partial_j v_j)$}, while the structures
 {$\phi' \partial^{2} (\partial_j v_j)$}, etc. are forbidden. Another
consequence is UV finiteness of the 1-irreducible function  {$\langle \phi'v_iv_j \rangle$} with
$\delta_{\Gamma}=2$. Each diagram
of this function contains the factor $c_{0}^{2}$, which forbids the
counterterms of the form  {$\phi' (\partial_i v_i)(\partial_j v_j)$},~etc., while the
remaining structure $c_{0}^{2}  \phi' v^{2}$ is forbidden by the Galilean
symmetry.

\item A crucial observation refers to the function  {$\left\langle v_i'v_j'\right\rangle$}: the corresponding index of 
divergence reads $\delta_\Gamma=-d+4$, therefore,
it becomes UV divergent in $d=2, 3, \text{and}\ 4$ and requires a presence of specific counterterms. 
For the physical case $d=3$ ($\delta_\Gamma=1$) it is impossible to construct a scalar counterterm containing two vector fields and one 
derivative\footnote{For the same reason the diagram  {$\left\langle v_i'v_j'v_k\right\rangle$} with $\delta_\Gamma=3-d$ does not diverge at $d=3$.}, so 
the only possible way is to include the UV cutoff $\Lambda$ in the counterterm. 
 Such counterterms do not involve poles in $y$ (or $\eps=4-d$, see later) and, therefore, they are lost if the calculations are performed using the formal rules of dimensional regularization and 
do not affect critical behavior. 
The situation is similar to the well-known $\phi^4$ model, in which 
such counterterms leads to the shift of the parameter $\tau=T-T_c$, the deviation of the
temperature from its critical value, which in theory of critical phenomena is 
an analogue of the mass term: $\tau_0\to\tau_0+g_0\Lambda^2$. 
The difference is that 
in our model there is no local term in~\eqref{power}, so this term should appear, with its own constant; see 
expression~\eqref{eq:correl2}. But if $d<4$ the new constant $g_{20}$ is not dimensionless and, therefore, 
does not give rise to the additional beta function in the RG equations\footnote{Detailed discussion of the similar situation in the RG analysis of
the helical magnetohydrodynamic (MHD) turbulence can be found in the monographs~\cite{Vasiliev}, Secs.~6.16 and 6.17, and~\cite{turbo}, Sec.~3.9.}.
This is why one can use the dimensional regularization in the analysis of the fixed points and in the calculation of critical exponents.

This means that in special space dimension $d=3$ the renormalization group analysis is simplified and does not catch features associated
with this divergence, and the results of~\cite{ANU97,AK14} should be treated as preliminary. 
A more fruitful approach
is to study our
model at $d=2$ or $d=4$, which may allow us to find new scaling regimes that can be applied to the physical value $d=3$.
\end{enumerate}

In this work we analyze the model~\eqref{eq:action} 
in the vicinity of the spatial dimension $4$, which requires to take into account only one
additional divergent function  {$\left\langle v_i'v_j'\right\rangle$}. For this reason we have modified the kernel
function $\widetilde{D}_{ik}({\bm k})$ [see~\eqref{power} and~\eqref{eq:correl2}] and introduced  the second coupling constant $g_{20}$ and
$\varepsilon=4-d$, which together with $y$ {plays} the role of an expansion parameter. To explore this model at $d=2$ one should consider four
new divergent functions, namely,  {$\left\langle v_i'v_j'\right\rangle$} with $\delta_\Gamma=2$, the functions
  {$\left\langle v_i'v_j'v_k'\right\rangle$} and  {$\left\langle v_i'v_j'v_k\right\rangle$} with $\delta_\Gamma=1$, and
 the function  {$\left\langle v_i'v_j'v_k'v_l'\right\rangle$} with $\delta_\Gamma=0$, so it is  a much more complicated task and a possible problem for the future studies{~\cite{Tomas}}.

Using all these considerations one can {show} that all the UV divergences in
the model~(\ref{eq:action}) {near} $d=4$ {can be} removed by the counterterms of the form
\begin{align}
& {v_{i}'\partial^{2} v_{i}},  &v_{i}'& \partial_{i} \partial_{j}v_{j}, 
&v_{i}'&\partial_{i}\phi, \nonumber \\ 
&c_{0}^{2} \phi'\partial_{i}v_{i}, 
&\phi'&\partial^{2} \phi,  & {v_i'}& {v_i'},
\label{counter}
\end{align}
which are already included
in the extended action functional~(\ref{eq:action}) with $v_{0}>0$.
Now the poles can be eliminated by multiplicative renormalization of the parameters $g_{10},g_{20},\nu_0,u_0,v_0,c_0$ and
 the fields $\phi$ and $\phi'$:
\begin{align}
g_{10} & =  g_1 \mu^y Z_{g_1}, &u_{0}& = u Z_{u}, &\nu_0& = \nu Z_{\nu},
\nonumber \\
g_{20} & =  g_2 \mu^\varepsilon Z_{g_2}, &v_{0}& = v Z_{v}, &c_{0}& = c Z_{c}.
\label{mult}
\end{align}
Here, $\mu$ is the scale-setting parameter (additional free parameter
of the renormalized theory) in the MS scheme, 
the parameters $g_1,g_2,\nu,u,v$, and $c$ are renormalized analogs of the bare
parameters (without subscript ``$0$''), $Z_i,i\in\{g_1,g_2,u,v,\nu,c\}$,
are the renormalization constants, which depend only on the completely
dimensionless parameters $g_1,g_2,u,v,\alpha,d,y$ and $\varepsilon$. 
The fields $\phi$ and $\phi'$ are renormalized in the following way:
\begin{eqnarray}
\phi \to Z_{\phi}\phi, \quad \phi' \to Z_{\phi'}\phi'.
\end{eqnarray}
The non-local part of 
the function $D_{ik}$
does not require the renormalization, so it can be expressed
in renormalized parameters using the relation $g_{10}\nu_0^{3} =
g_1\nu^3\mu^y$, see~\eqref{Dr} below. 
The parameters $m$ and $\alpha$ from the correlation function~(\ref{ff}) are
not renormalized: $Z_{m}=Z_{\alpha}=1$. Due to the absence of renormalization
of the term  {$v_i'\nabla_{t}v_i$} no renormalization of the fields
 {$v_i$} and  {$v_i'$} is needed: $Z_{v}=Z_{v'}=1$.

Hence, the renormalized action functional has the form
\begin{widetext}
\begin{eqnarray}
{\cal S}^{R}_{ {\bm v} }(\Phi) &=& \frac{1}{2} v_{i}'D^{R}_{ij}  v_{j}' +
v_{i}' \biggl[ -\nabla_{t} v_{i} +
Z_{1} \nu (\delta_{ij}\partial^{2}-\partial_{i}\partial_{j}) v_{j} +
Z_{2} u \nu\partial_{i}\partial_{j} v_{j} - Z_{4}\partial_{i} \phi \biggr] +
\nonumber \\
&+& \phi' \left[ -\nabla_{t}\phi  + Z_{3}v \nu \partial^{2} \phi -
Z_{5} c^{2}(\partial_{i}v_{i}) \right],
\label{Raction}
\end{eqnarray}
\end{widetext}
where 
\begin{equation}
D^R_{ij}=g_{1} \mu^y\nu^3  p^{4-d-y} \biggl\{
  P_{ij}({\bm p}) + \alpha Q_{ij}({\bm p})
  \biggl\} 
  + Z_6 g_{2}\mu^\varepsilon \nu^3 \delta_{ij}.
\label{Dr}
\end{equation}
In comparison to the case $d=3$, there additional renormalization constant is needed, namely $Z_6$.

\section{Renormalization of the model}\label{sec:Renorm}

\subsection{Perturbation expansion for the 1-irreducible Green functions}
\label{Zs}

Let us consider the generating functional  {$\Gamma(\Phi)$} of the {1-irreducible} Green functions. 
According to Eq. (\ref{eq:expansion1}), $\Gamma(\Phi)$ can be written using the Legendre transform in the following form
\begin{equation}
   {\Gamma(\Phi) =  \S_{{{\bm v}}}(\Phi) + \widetilde{\Gamma}(\Phi)},
  \label{eq:expansion}
\end{equation}
where for the functional arguments we have used the same symbols
$ {\Phi}=\left\{ { v_i}, { v_j'},\phi,\phi' \right\}$ as for the corresponding random fields.
Here, ${\S}_{{{\bm v}}}( {\Phi})$ is the action functional~(\ref{eq:action}) and
$\widetilde{\Gamma}( {\Phi})$ is the sum of all the 1-irreducible diagrams with loops.
Hence, in the one-loop approximation, the expressions for the {1-irreducible} Green
functions that require UV renormalization take the form
\begin{widetext}
\begin{align}
   \Gamma_{v'v} & = i\omega -
 (\delta_{ij}p^2 - p_i p_j)Z_1 \nu - p_i p_j Z_2 u\nu +  \raisebox{-1.ex}{ \includegraphics[width=2.5truecm]{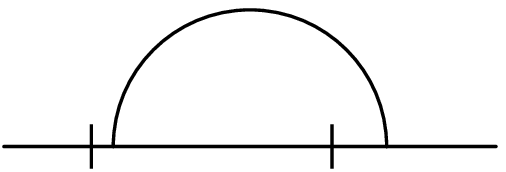}},
 \label{eq:vsv}\\
     \Gamma_{\phi \phi'} & = i\omega - p^2 Z_3 v\nu + \raisebox{-1ex}{ \includegraphics[width=2.5truecm]{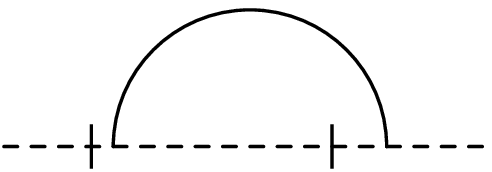}},
    \label{eq:phiphis}\\
    \Gamma_{v'\phi} & = -iZ_4 p_i +\raisebox{-1ex}{ \includegraphics[width=2.truecm]{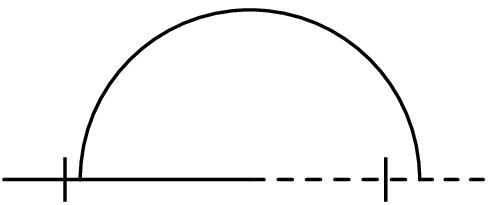}},
    \label{eq:vsphi}   \\
    \Gamma_{\phi' v} & =  {-iZ_5 p_j c^2} +\raisebox{-1ex}{ \includegraphics[width=2.5truecm]{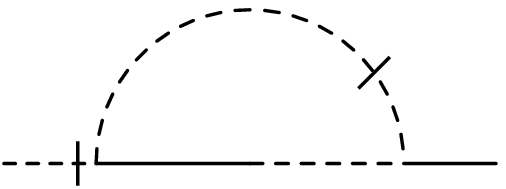}}
      +\raisebox{-1ex}{ \includegraphics[width=2.5truecm]{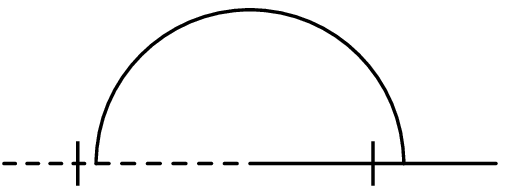}}
      +\raisebox{-1ex}{ \includegraphics[width=2.5truecm]{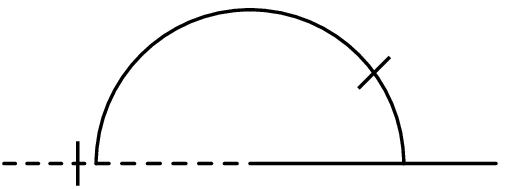}},
      \label{eq:phisv} \\
    \Gamma_{v'v'} & =  g_{1}\mu^y \nu^3 p^{4-d-y} \biggl\{
  P_{ij}({\bm p}) + \alpha Q_{ij}({\bm p})
  \biggl\} 
  + Z_6 g_{2} \mu^\varepsilon\nu^3 \delta_{ij} + \frac{1}{2}\raisebox{-1.0ex}{ \includegraphics[width=2.5truecm]{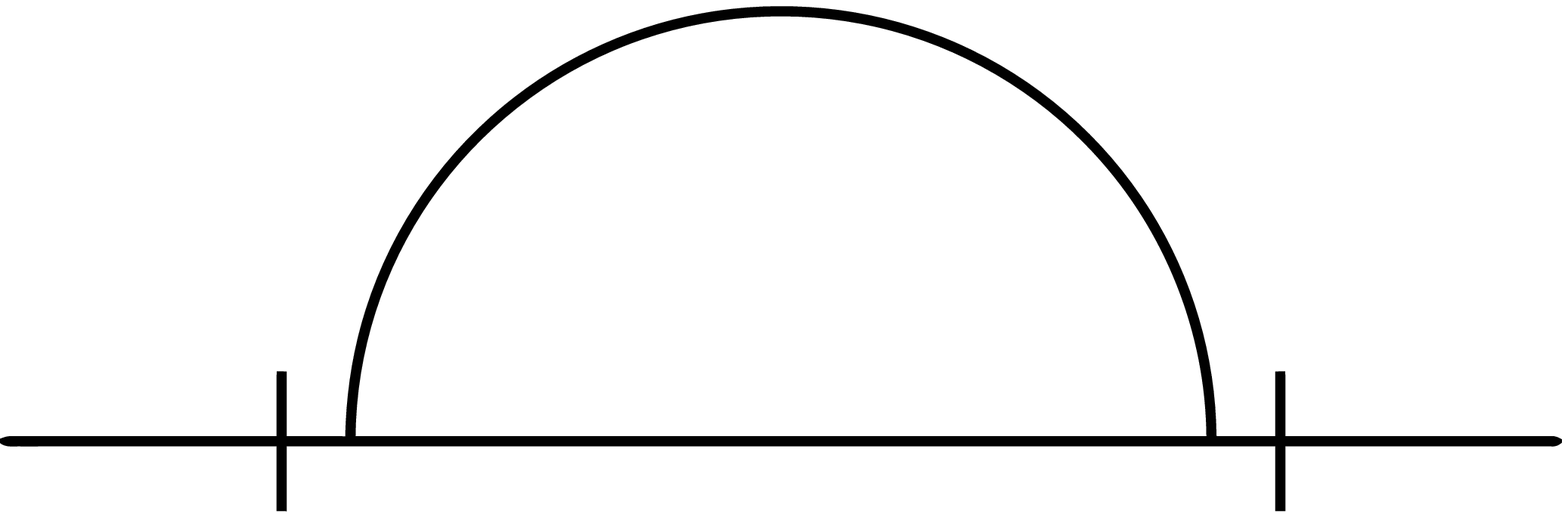}},
  \label{eq:vsvs}
\end{align}
\end{widetext}
where ${\bm p}$ stands for the corresponding external momentum. 
The factor $1/2$ in front of the diagram in~(\ref{eq:vsvs}) is the symmetry coefficient of the graph; for all the other graphs
the symmetry coefficients are equal to $1$.

From a direct comparison of the relations between renormalized parameters it is straightforward to show that
the renormalization constants in~(\ref{mult}) 
and~\eqref{eq:vsv}~-- \eqref{eq:vsvs} are related as follows
\begin{align}
   Z_\nu &= Z_1, &Z_{g_1}& = Z_1^{-3}, &Z_c& = (Z_4 Z_5)^{1/2}, \nonumber\\
   Z_\phi& = Z_4, &Z_{\phi'} &= Z_4^{-1}, &Z_v& = Z_3 Z_{1}^{-1}, \nonumber\\
   Z_u& = Z_2 Z_1^{-1}, &Z_{g_{2}}& = Z_6 Z_1^{-3}.
\label{eq:RGconst2}
\end{align}
The renormalization constants are derived from the requirement that the
Green functions of the renormalized model~(\ref{Raction}), when expressed
in renormalized variables, be UV finite.

\subsection{Renormalization constants}
 {All diagram calculations are performed using dimensional regularization and the MS scheme, and can be found in Appendix~\ref{app1}. 
All the diagrams are calculated in the arbitrary space dimension $d$, and only the poles in $y$ and $\eps=4-d$ are presented in the results.} 

 {The renormalization constants of the fields $\phi$ and $\phi'$ and the physical parameters of the system calculated from the diagrams and expressions~\eqref{eq:vsv}~-- \eqref{eq:vsvs} and~\eqref{eq:RGconst2} are:}
\begin{align}
  Z_\nu & = 1 + A \biggl( \frac{g_1}{y} + \frac{g_2}{\eps}\biggl)    +
   B\biggl(\alpha  \frac{g_1}{y} + \frac{g_2}{\eps}\biggl); \nonumber \\
  Z_{u} & = 1 + 
  (C-A) \biggl( \frac{g_1}{y} + \frac{g_2}{\eps}\biggl)
  -B  \biggl( \alpha\frac{g_1}{y} + \frac{g_2}{\eps}\biggl); \nonumber \\
  Z_{v} & = 1 - (A+D) \biggl( \frac{g_1}{y} + \frac{g_2}{\eps}\biggl) \nonumber \\
  &- (B+E) \biggl( \alpha \frac{g_1}{y} + \frac{g_2}{\eps}\biggl); \nonumber \\
  Z_{c} & = 1 + \frac{1}{2}F \biggl( \frac{ g_1}{y} + \frac{g_2}{\eps}\biggl) ;\nonumber \\
  Z_{\phi} & = 1 + F \biggl( \frac{ g_1}{y} + \frac{g_2}{\eps}\biggl);\nonumber  \\
  Z_{\phi'} & =1 - F \biggl( \frac{ g_1}{y} + \frac{g_2}{\eps}\biggl); \nonumber \\
  Z_{g_1} & = 1 - 3 A \biggl( \frac{g_1}{y} + \frac{g_2}{\eps}\biggl) -3B \biggl( \alpha\frac{ g_1}{y} + \frac{g_2}{\eps}\biggl); \nonumber \\
  Z_{g_2} & = 1 - 3 A \biggl( \frac{g_1}{y} + \frac{g_2}{\eps}\biggl) -3B \biggl( \alpha\frac{ g_1}{y} + \frac{g_2}{\eps}\biggl) \nonumber \\
  &-G\biggl[ \alpha \frac{g_1^2}{g_2(2y-\eps)}+ (1+\alpha)\frac{g_1}{y} + \frac{g_2}{\eps}  \biggl],
\label{Zpar}
\end{align}
where $A,\ldots,F$ are the coefficients {from} the renormalization constants $Z_1$~-- $Z_6$ (see Appendix~\ref{app1}):
\begin{align}
\label{ZA}
  A & =-\frac{d(d-1)u^2+2u(d^2+d-4)+d(d+3)}{4d(d+2)(u+1)^2};\nonumber \\
  B & =-\frac{u-1}{2du(1+u)^2};\nonumber \\
  C & = -(d-1)\frac{u^2(d-1) +u(d+4) + 1}{2d(d+2)u(u+1)^2};\nonumber \\
  D & = \frac{d-1}{2dv(v+1)}; \nonumber \\
  E & = \frac{u-v}{2dvu(u+v)^2}; \nonumber \\
  F & = \frac{d-1}{2 d(u+1)(v+1)};\nonumber \\
  G & = \frac{d-1}{2d u(u+1)}.
\end{align}
 {The expressions~\eqref{Zpar} contain all renormalization constants needed to renormalize our model {near} $d=4$.}

\subsection{RG equations and functions}
\label{sec:rgrg}

The relation between the initial and renormalized action functionals $\S(\Phi,e_{0})= \S^{R}(Z_\Phi\Phi,e,\mu)$ (where $e_{0}$
is the complete set of bare parameters and $e$ is the set of their renormalized
counterparts) yields the fundamental RG differential equation:
\begin{equation}
\biggl\{ {\cal D}_{RG} + N_{\phi}\gamma_{\phi} +
N_{\phi'}\gamma_{\phi'} \biggr\} \,G^{R}(e,\mu,\dots) = 0,
\label{RG1}
\end{equation}
where $G =\langle \Phi\cdots\Phi\rangle$ is a correlation function of the fields~$\Phi$;
$N_{\phi}$ and $N_{\phi'}$ are the counts of normalization-requiring fields $\phi$ and $\phi'$, respectively, which are the inputs to
$G$; 
the ellipsis in expression~\eqref{RG1} stands for the other arguments of $G$ (spatial
and time variables, etc.).
${\cal D}_{RG}$ is the operation $\widetilde{\cal D}_{\mu}$
expressed in the renormalized variables and
$\widetilde{\cal D}_{\mu}$
is the differential operation $\mu\partial_{\mu}$ for fixed
$e_{0}$. For the present model it takes the form
\begin{equation}
{\cal D}_{RG}= {\cal D}_{\mu} + \beta_{g_1}\partial_{g_1} + \beta_{g_2}\partial_{g_2} +
\beta_{u}\partial_{u} + \beta_{v}\partial_{v}
- \gamma_{\nu}{\cal D}_{\nu}- \gamma_{c}{\cal D}_{c} .
\label{RG2}
\end{equation}
Here, we have denoted ${\cal D}_{x} \equiv x\partial_{x}$ for any variable $x$.
The anomalous dimension $\gamma_{F}$ of some quantity $F$
(a field or a parameter) is defined as
\begin{equation}
\gamma_{F}= Z_{F}^{-1} \widetilde{\cal D}_{\mu} Z_{F} =
\widetilde{\cal D}_{\mu} \ln Z_F ,
\label{RGF1}
\end{equation}
and the $\beta$ functions for the four dimensionless coupling
constants $g_1$, $g_2$, $u$ and $v$,
which express the flows of parameters under
the RG transformation, are $\beta_{g} = \widetilde{\cal D}_{\mu} g$. 
Together with~(\ref{mult}) this yields
\begin{align}
\beta_{g_1} & =  g_1\,(-y-\gamma_{g_1}),
&\beta_{g_2} & =  g_2\,(-\varepsilon-\gamma_{g_2}),
\nonumber \\
\beta_{u} & =  -u\gamma_{u},
&\beta_{v} & = -v\gamma_{v}.
\label{betagw}
\end{align}
From the definitions and explicit expressions~\eqref{Zpar},~\eqref{ZA} one finds in the one-loop approximation 
(i.e., with corrections of orders $g_1^2$, $g_2^2$, $g_1g_2$ and higher) in $d=4$:

\begin{widetext}
\begin{align}
\label{gammanu}
  \gamma_\nu &  = g_1\frac{3u^2+8u+7}{24(u+1)^2}+  \alpha g_1\frac{u-1}{8u(u+1)^2}  +g_2\frac{3u^3+8u^2+10u-3}{24u(u+1)^2}; \\
  \gamma_{u} &   = -\frac{u-1}{48u(u+1)^2}\biggl[ g_1(6u^2+13u+3) + 6\alpha g_1 + g_2(6u^2+13u+9)  \biggl]; \\
  \gamma_{v} &  = \frac{g_1}{24}\biggl[ -\frac{3u^2+8u+7}{(u+1)^2}+\frac{9}{v(v+1)}\biggl] - \alpha g_1\frac{v-1}{8u(u+1)^2v(u+v)^2}
  \biggl[u^3+2u^2(v+1)-v(v+1)+u(v^2-v+1)\biggl] \nonumber \\
  &+\frac{g_2}{24}\biggl[-\frac{3(u-1)}{u(u+1)^2}-\frac{3u^2+8u+7}{(u+1)^2}+\frac{3(u-v)}{uv(u+v)^2}+\frac{9}{v(v+1)}\biggl];  \\
  \gamma_{c} &  = -\frac{3}{16(u+1)(v+1)}(g_1+g_2); \\
  \gamma_{\phi} & = -\frac{3}{8(u+1)(v+1)}(g_1+g_2); \\
  \gamma_{\phi'} & = \frac{3}{8(u+1)(v+1)}(g_1+g_2); \\
  \gamma_{g_1}&  = -g_1\frac{3u^2+8u+7}{8(u+1)^2}-\alpha g_1\frac{3(u-1)}{8u(u+1)^2} -g_2\frac{3u^3+8u^2+10u-3}{8u(u+1)^2}; \\
  \gamma_{g_2} & = \frac{1}{8u(u+1)^2}\biggl[ -g_1(3u^3+8u^2+4u-3) - g_2(3u^3+8u^2+7u-6) +3\frac{\alpha g_1}{g_2} [(u+1)g_1+2g_2] \biggl].
\label{gammag2}
\end{align}
\end{widetext}

This means that from the expressions~\eqref{betagw} and~\eqref{gammanu}~-- \eqref{gammag2} all {the functions entering
the differential operator}~\eqref{RG2} are known,  and, therefore, {now we may consider how {this differential 
operator} acts on different Green functions}. We do not
include the dimensionless parameter $\alpha$ into the list of coupling constants, because it is not renormalized ($Z_{\alpha}=1$) and the
corresponding function $\beta_{\alpha}$ vanishes identically.
Thus, the RG equations do not impose restrictions on the value of $\alpha$, and $\alpha$ remains a free parameter of the model.

\section{Renormalization group and critical scaling} \label{sec:RC}

\subsection{RG functions and IR attractive fixed points}

From the analysis of the RG equation~\eqref{RG1} it follows that 
the large scale behavior with respect to spatial and time scales is
governed by the IR attractive (``stable'') fixed points $g^*\equiv\{g_1^*,g_2^*,u^*,v^*\}$,
 {whose} coordinates are found from the conditions~\cite{Vasiliev,Zinn,Amit}:
\begin{align}
  &\beta_{g_1} (g^{*}) = \beta_{g_2} (g^{*})= \beta_{u}
  (g^{*}) = \beta_{v} (g^{*}) = 0.
  \label{eq:gen_beta}
\end{align}
Consider a set of invariant couplings $\overline{g}_i = \overline{g}_i(s,g)$
with the initial data $\overline{g}_i|_{s=1} = g_i$. Here, $s=k/\mu$ 
and IR asymptotic behavior (i.e., behavior at large distances) corresponds
to the limit $s\rightarrow 0$. An evolution of invariant couplings is described by
the set of flow equations
\begin{equation}
  \mathcal{D}_s \overline{g}_i = \beta_i(\overline{g}_j),
  \label{eq:invariant_chrg}
\end{equation}
whose solution as $s\to0$ behaves approximately like
\begin{equation}
  \overline{g}_i(s,g^*) \cong g^*+const\times s^{\omega_i},
  \label{Asym}
\end{equation}
where $\left\{\omega_i\right\}$ is the set of eigenvalues of the matrix 
\begin{equation}
\Omega_{ij}=\partial\beta_{i}/\partial g_{j}|_{g=g_{*}}.
\label{Omega}
\end{equation}
The existence of IR {attractive} solutions of the RG equations leads
to the existence of the scaling behavior of Green functions. 
From~\eqref{Asym} it follows that the type of the fixed point is determined by the matrix~\eqref{Omega}:
for the IR {attractive} fixed points the matrix $\Omega$ has to be positive definite.

In contrast to the three dimensional case, where the analysis of the expressions like~\eqref{betagw} and~\eqref{gammanu}~-- \eqref{gammag2}
has shown that in the physical range of parameters $g_1, u, v, \alpha>0$ 
{there exist only two IR attractive fixed points, one trivial (Gaussian) fixed point and one non-trivial~\cite{ANU97,AK14},} at $d=4$ 
situation is more intriguing:
a direct analysis of the system of equations~(\ref{eq:gen_beta}) reveals the existence of three
IR attractive fixed points: a trivial free fixed point (\fp{I}) and two non-trivial fixed points (\fp{II} and \fp{III}). 

The point \fp{I}, for 
which all interactions are irrelevant and no scaling and universality is expected, has the coordinates
\begin{equation}
  g_1^* = 0, \quad g_2^* = 0, 
  \label{eq:trivialFP}
\end{equation}
whereas the coordinates for couplings $u^*$ and $v^*$ {are arbitrary}.
The corresponding eigenvalues of the matrix $\Omega_{ij}$ are
\begin{equation}
\label{L1}
   \lambda_1 = 0,\quad 
   \lambda_2 = 0,\quad
   \lambda_3 = -\eps,\quad 
   \lambda_4 = -y.
\end{equation}
Though trivial, \fp{I} is necessary for the correct use of the perturbative renormalization group. From~\eqref{L1} it follows that \fp{I} 
is IR attractive for negative values of $y$ and $\varepsilon$. 
Expressions~\eqref{eq:trivialFP} and~\eqref{L1} imply that in the 
four-dimensional space of coupling constants $\left\{g_1,g_2,u,v\right\}$ this fixed point is a ``point'' only in two
dimensions $\left\{g_1,g_2\right\}$, and in the four-dimensional space of all couplings it is a two-dimensional plane.
Zero eigenvalues $\lambda_1$ and $\lambda_2$ correspond to zero velocity along this plane, perpendicular to the plane $\left(g_1,g_2\right)$.

For the second fixed point, \fp{II}, $g_1^* = 0$ while $g_2^* \neq 0$. Therefore, this scaling regime is called ``local''~[see~\eqref{eq:correl2}].
Its coordinates are
\begin{equation}
  g_1^* = 0, \quad g_2^* = \frac{8\eps}{3}, \quad u^* = 1, \quad v^* = 1.
\label{gfp2}
\end{equation} 
The eigenvalues of the matrix $\Omega_{ij}$ are
\begin{equation}
\label{Lambda2}
   \lambda_1 = \frac{7\eps}{18},\quad 
   \lambda_2 = \frac{5\eps}{6},\quad
   \lambda_3 = \eps,\quad 
   \lambda_4 = \frac{3\eps-2y}{2}.
\end{equation}
Thus, \fp{II} is IR attractive in the region satisfying the inequalities $\varepsilon>0$ and $y<3\varepsilon/2$ 
and it is a node attractor, see discussion below.

For the last fixed point, \fp{III}, both the non-local and local parts of the random force are relevant:
\begin{align}
  g_1^* &= \frac{16y(2y-3\eps)}{9[y(\alpha+2)-3\eps]}, 
  &g_2^*& =\frac{16\alpha y^2}{9[y(\alpha+2)-3\eps]}, \nonumber\\
  u^* &= 1, &v^*& = 1.  
\label{gfp3}
\end{align}
The {corresponding} eigenvalues of the matrix $\Omega_{ij}$ are
\begin{align}
\label{Lambda3}
   \lambda_1 &= \frac{y[2y(10\alpha+11)-3\eps(3\alpha+11)]}{54[y(\alpha+2)-3\eps]};\nonumber \\
   \lambda_2 &=\frac{y[2y(2\alpha+3)-\eps(\alpha+9)]}{6[y(\alpha+2)-3\eps]}; \nonumber \\
   \lambda_{3,4} &= \frac{A\pm \sqrt{B}}{C},
\end{align}
where the constants $A$, $B$, and $C$ are given by
\begin{align}
\label{ABC}
  A & = -27 \eps^3+9 (\alpha+9) \eps^2 y-9 (3\alpha+8) \eps y^2 \nonumber\\
  & +2y^3(\alpha^2+7\alpha+10);\nonumber \\
  B & =[-3 \eps + (\alpha+2) y]^2 [81 \eps^4-54 \eps^3 y-9 (20\alpha+3) \eps^2 y^2\nonumber\\
  &+12 (3 \alpha^2+17 \alpha+1) \eps y^3 -4 (5\alpha^2+14 \alpha-1) y^4]; \nonumber \\
  C & = 6 [-3 \eps+(\alpha+2) y]^2.
\end{align} 
Taking into account that in the physical range the couplings $g_1$ and $g_2$ must be positive, it follows from the explicit 
form of the eigenvalues $\lambda_1\ldots\lambda_4$ 
that the point \fp{III} is IR attractive when $y>0$ and $y>3\eps/2$. 

Furthermore, from the explicit form of the $\beta$ functions $\beta_u$ and $\beta_v$ it readily
follows that the $4\times4$ matrix $\Omega_{ij}$ decomposes to three blocks, the first two are $1\times1$ and the third is $2\times2$.
Two $1\times1$ blocks are determined by the eigenvalues $\lambda_1=\partial\beta_u/\partial u|_{g=g*}$ and 
$\lambda_2=\partial\beta_v/\partial v|_{g=g*}$, see~\eqref{Lambda3}.
The remaining block, which needs to be diagonalized, is a $2\times2$ matrix, denoted 
$\widetilde{\Omega}_{ij}$. This decomposition opens another
opportunity to analyze whether \fp{III} is IR attractive: the matrix $\widetilde{\Omega}_{ij}$ is positive definite if 
and only if both $\text{Tr}\,\widetilde{\Omega}_{ij}>0$ and $\text{Det}\,\widetilde{\Omega}_{ij}>0$. In our case 
\begin{align}
\label{Trdet}
\text{Tr}\,\widetilde{\Omega}_{ij}  &=\frac{9\eps^2-12\eps y+2(\alpha+5)y^2}{3[(\alpha+2)y-3\eps]};\nonumber\\
\text{Det}\,\widetilde{\Omega}_{ij} &=y\left(\frac{2}{3}y-\eps\right).
\end{align}
From~\eqref{Trdet} it follows that the matrix $\widetilde{\Omega}_{ij}$ is positive in the region $y>0$ and $y>3\eps/2$. This approach is
simpler than direct analysis of the expressions~\eqref{Lambda3}~-- \eqref{ABC}, but does not distinguish a simple node attractor from a more complicated spiral attractor. The latter is a consequence of a non-zero imaginary part in the eigenvalues of the matrix $\Omega_{ij}$. 

The ability to determine whether an IR attractive fixed point corresponds to
a node or a spiral attractor is an advantage of the double $y$ and $\varepsilon$ expansion near $d=4$. Indeed, in 
the case of a simplified analysis near $d=3$~\cite{ANU97,AK14}, $\Omega_{ij}$ is a $3\times3$ matrix and its
eigenvalues\footnote{Note, that in previous study~\cite{AK14} there are misprints in the expressions~(2.25) and~(2.28) for the constant $Z_3$ and
function $A(d)$, which enter into expressions for $\beta$ functions $\beta_g$, $\beta_u$, and $\beta_v$. The correct expressions read
$$
Z_3 = 1 - \frac{\hat{g}}{y} \frac{d-1}{2dv(v+1)} -\frac{\alpha \hat{g}}{y} \frac{u-v}{2duv(u+v)^2}
$$
and
$$
A =\frac{-d(d-1)u^{2} - 2(d^{2}+d-4)u - d(d+3)}{4d(d+2)(1+u)^{2}} +
\frac{\alpha(1-u)}{2du(1+u)^{2}}.
$$
} are
\begin{align}
\label{d3ein}
 {
\lambda_1=y; \quad
\lambda_2=\frac{\alpha+6}{12}y; \quad
\lambda_3=\frac{5\alpha+12}{96}y}. 
\end{align}
From expressions~\eqref{d3ein} it follows that all the eigenvalues ($\lambda_1, \lambda_2, \lambda_3$) are real. 
Therefore, the fixed point is a node attractor. 
Nevertheless, this point may be a spiral attractor at large values of $y$, 
but we are not able to investigate this question near $d=3$. 

However, we can perform this analysis near $d=4$.
The quantity $B$ [see~\eqref{ABC}] is a fourth order polynomial in the exponent $y$. From its analysis it follows that 
if $\alpha\leq\frac{1}{5}(-7+3\sqrt{6})\approx0.07$, which is a positive root of the equation $5\alpha^2+14\alpha-1=0$, the expression
$B$ is strictly positive in the region $y>0$, $y>3\eps/2$.
That is, in this case \fp{III} is a node attractor for all permissible values of $y$ and $\eps$. 
If $\alpha>\frac{1}{5}(-7+3\sqrt{6})$, equation $B(y)=0$ has one root $R(\alpha,\eps)$, that is larger than $3\eps/2$. 
Thus, if $3\varepsilon/2<y<R(\alpha,\varepsilon)$, FPIII is a node attractor; 
whereas if $y>R(\alpha,\varepsilon)$, the eigenvalues $\lambda_{3}$ and $\lambda_4$ contain non-zero imaginary parts and FPIII is a spiral attractor.
An abrupt change from a node to a spiral attractor in the region
$y>R(\alpha,\varepsilon)$ near the point $\alpha=\frac{1}{5}(-7+3\sqrt{6})$ (especially at $\eps<0$) looks intriguing, but may be an
artifact of the one-loop approximation we used. 

\begin{figure}[t]
\center
\includegraphics[width=0.43\textwidth,clip]{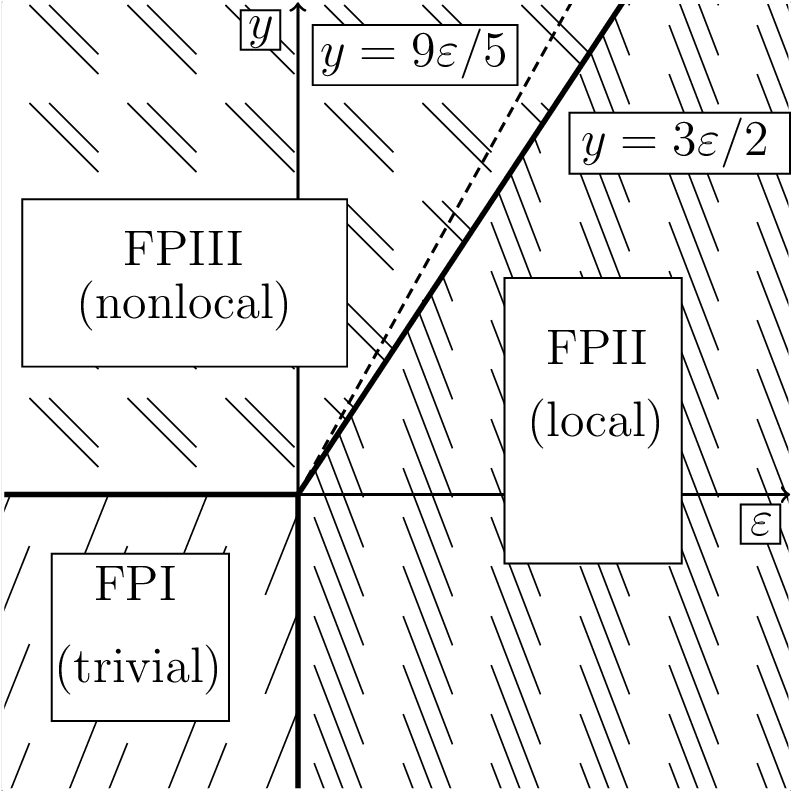}
\caption{Domains of IR stability of the fixed points for the model~(\protect\ref{eq:action}) 
in the plane $(y,\eps)$.}
\label{fig:Plain}
\end{figure}

The explicit expression for $R(\alpha,\eps)$, being a solution of the fourth degree polynomial equation, is rather 
cumbersome and, therefore, is omitted here. 
We have computed that
\begin{equation}
\label{lim}
\lim_{\alpha\to\infty} R(\alpha,\varepsilon) = \frac{9\varepsilon}{5},
\end{equation}
which is in agreement with~\eqref{AIW}, obtained as a result of the analysis performed directly at $\alpha=\infty$. This means that the crossover between node and spiral attractors 
takes place along the line $R(\alpha,\varepsilon)$, which is vertical at $\alpha$ near $\frac{1}{5}(-7+3\sqrt{6})$ and 
rotates clockwise toward the line $R(\infty,\varepsilon) = 9\varepsilon/5$ as $\alpha\to\infty$ (see Fig.~\ref{fig:Plain}).

At the real values of the parameters $y=4$ and $\varepsilon=1$ \fp{III} is a spiral
attractor for $\alpha>\frac{5}{176} (-26 + 9 \sqrt{15})\approx0.251$.

To complete the analysis of the fixed point structure, infinite fixed point values $u\to\infty$ and $v\to\infty$ have to 
be considered {as well}.
Since $u$ may be interpreted as longitudinal viscosity, from the physical point of view this case corresponds to the limit
$c\to\infty$. Here, the velocity fields $ {v_i}$ and $ {v_i'}$ are purely transverse and the
scalar fields $\phi$ and $\phi'$ are effectively decoupled from them; see the explicit expressions for propagators~\eqref{lines}.
By introducing a new variable $f=1/u$ with its $\beta$ function
\begin{equation}
\label{F}
\beta_f=\widetilde{{\cal D}}_\mu f=-f^2\beta_u,
\end{equation}
one obtains the following set of $\beta$ functions at $f=0$:
\begin{align}
\label{betaF}
\beta_{g_1}&=\frac{1}{8}g_1(3g_1+3g_2-8y); \nonumber \\
\beta_{g_2}&=\frac{1}{8}g_2(3g_1+3g_2-8\varepsilon); \nonumber \\
\beta_{v}&=\frac{1}{8}(g_1+g_2)\frac{v^2+v-3}{v+1}.
\end{align}
From~\eqref{F} and~\eqref{betaF} it follows that there are
two non-trivial fixed points for $f^{*}=0$:
\begin{align}
\label{11}
g_1^{*} & = 0, &g_2^{*}& = 8\varepsilon/3, &v^{*}& =\frac{1}{2}(-1+\sqrt{13});\\
g_1^{*} & = 8y/3, &g_2^{*}& = 0, &v^{*}& = \frac{1}{2}(-1+\sqrt{13}).
\label{12}
\end{align}
However, two of the four eigenvalues have opposite signs at any values of $y$ and $\varepsilon$, namely
\begin{equation}
\lambda_1=-y/3, \quad \lambda_2=\frac{2(13+\sqrt{13})}{3(1+\sqrt{13})^2}y
\end{equation}
for the fixed point given by Eq.~\eqref{11}, whereas
\begin{equation}
\lambda_1=-\varepsilon/3, \quad \lambda_2=\varepsilon
\end{equation}
for the fixed point given by Eq.~\eqref{12}.

Thus, both fixed point~\eqref{11} and~\eqref{12} are unstable (i.e., saddle points). This agrees with 
the observation that the leading-order correction in the Mach number to the incompressible scaling
regime destroys its stability~\cite{LM,KompOp,NV}.

In order to study the limit $v\rightarrow\infty$, let us create a new variable $t=1/v$ with $\beta$ function
\begin{equation}
\label{T}
\beta_t=\widetilde{{\cal D}}_\mu t=-t^2\beta_v.
\end{equation}
For  $t=0$ one obtains $\beta_t=0$. Since $\beta$ functions of the other coupling 
constants $g_1$, $g_2$, and $u$ are independent of $v$,  at $t=0$ we recognize the formerly obtained fixed points II and III. Thus, 
to investigate the IR attraction of these two points, one should only check the derivative $\partial\beta_t/\partial t$ at the fixed point $\left\{g^*, t=0\right\}$:
\begin{align}
\lambda_t&=-\varepsilon/2 \quad \text{for \fp{II}}; \nonumber \\
\lambda_t&=-y/3 \quad \text{for \fp{III}}.
\label{MashaTMP}
\end{align}
Comparing~\eqref{MashaTMP} with~\eqref{Lambda2} and~\eqref{Lambda3}, we find that 
in the limit $v\to\infty$ these two fixed points are saddle points as well.

In contrast to the direct analysis near three-dimensional case $d=3$ (see~\cite{AK14}), where non-trivial IR attractive fixed point
was valid for all $\alpha>0$, had finite limit at $\alpha\to\infty$, but was unstable at $\alpha=\infty$ (i.e., in the
case of a purely potential random force), under this analysis near $d=4$ the situation is much better.
Taking into account~\eqref{eq:correl2}, to study this limit we define new coupling constant $g_{1}'=g_{1}\alpha$, which is finite as
$\alpha\to\infty$; $g_{2}$ herewith does not change, i.e., $g_{2}'=g_{2}$. Hence, since $Z_\alpha=1$, a new $\beta$ function is
\begin{equation}
\beta_{g'_{1}}=\widetilde{{\cal D}}_\mu g_{1}'=\alpha\beta_{g_{1}},
\end{equation}
and the full set of $\beta$ functions is:
\begin{align}
\label{A}
 \beta_{g_{1}'}&=g_{1}'\biggl[-y + \frac{g_2 (3u^3+8u^2+10u-3)
  + 3g_1'(u-1)}{8u(u+1)^2}\biggl]; \nonumber \\
\beta_{g_{2}}&=g_{2}\biggl[-\varepsilon + \frac{g_2(3u^3+8u^2+7u-6)
-6g_1'}{8u(u+1)^2} \biggl]; \nonumber \\
\beta_{u}&=\frac{u-1}{48(u+1)^2}\left[6g_1'+g_2(6u^2+13u+9)\right]; \nonumber \\
\beta_{v}&=g_1'\frac{v-1}{8u(u+1)^2(u+v)^2} \nonumber \\
&\times\left[u^3+2u^2(v+1)-v(v+1)+u(v^2-v+1)\right]\nonumber \\
&-g_2\frac{v}{24}\left[\frac{3(1-u)}{u(u+1)^2}-\frac{7+8u+3u^2}{(u+1)^2}\right.\nonumber \\
&\left.+\frac{3(u-v)}{uv(u+v)^2}+\frac{9}{v(v+1)}\right].
\end{align} 

The solution {of} the system~\eqref{eq:gen_beta} in this case allows three IR attractive fixed points: a trivial one 
\begin{equation}
\label{FGPA}
  {g'}_1^* = 0, \quad g_2^* = 0, 
\end{equation}
with $u^*$ and $v^*$ undetermined, which is IR attractive when $y<0$ and $\varepsilon<0$;
a local one 
\begin{equation}
\label{FP2}
  {g'}_1^* = 0, \quad g_2^* = \frac{8\eps}{3}, \quad u^* = 1, \quad v^* = 1,
\end{equation}
which is IR attractive when $\varepsilon>0$ and $y<3\eps/2$; and a non-local one 
\begin{equation}
\label{FGPA2}
  {g'}_1^* = \frac{16}{9}(2y-3\eps), \quad g_2^* = \frac{16y}{9}, \quad u^* = 1, \quad v^* = 1,
\end{equation}
which is IR attractive when $y>0$ and  $y>3\eps/2$. The latter, being a non-local fixed point in the case of a purely potential
random force, can be obtained from expressions~\eqref{gfp3} in the limit $\alpha\to\infty$, taking together with the 
substitution $g_1=g'_1/\alpha$. Thus, in contrast to the simplified analysis near $d=3$~\cite{ANU97,AK14}, the 
analysis near $d=4$ provides a non-local fixed point~\eqref{gfp3}, which has a finite limit as $\alpha\to\infty$, corresponding to a 
longitudinal random force.

\begin{figure}[t]
\center
\includegraphics[width=.55\textwidth,clip]{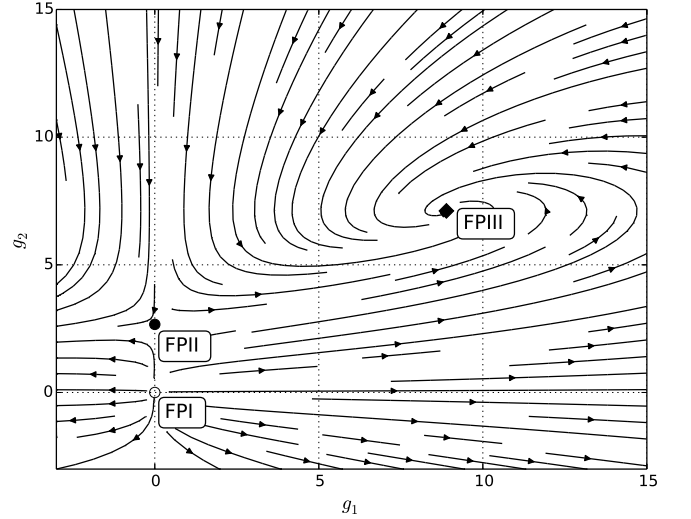}
\caption{RG flow diagram in the plane $(g_1,g_2)$ at 
$y=4$ and $\eps=1$; $\alpha=\infty$; $u=v=1$. Three fixed points \fp{I}, \fp{II}, and \fp{III} are marked by 
an empty circle, a filled circle, and a filled rhombus, respectively.}
\label{fig:RG}
\end{figure}

The eigenvalues of the matrix $\Omega_{ij}$ in this case equal to
\begin{align}
\label{AIW}
\lambda_1&=\frac{1}{6}(-\varepsilon+4y); &\lambda_2&=\frac{1}{54}(-9\varepsilon+20y); \nonumber\\
 \lambda_3&=\frac{1}{3}\biggl[y-\sqrt{(9\varepsilon-5y)y}\biggl]; &\lambda_4&=\frac{1}{3}\biggl[y+\sqrt{(9\varepsilon-5y)y}\biggl].
\end{align}

None of these eigenvalues contain an
imaginary part if $y<9\varepsilon/5$. Thus, a node attractor is realized
if $3\varepsilon/2<y<9\varepsilon/5$ and a spiral attractor is realized if $y>9\varepsilon/5$.

An RG flow diagram in the plane $(g_1,g_2)$ for $u=v=1$, $\alpha=\infty$, and real values of the parameters $y=4$ and $\eps=1$ is shown in Fig.~\ref{fig:RG}. 
The coordinates of three fixed points \fp{I}, \fp{II}, and \fp{III} are given by expressions~\eqref{FGPA}~-- \eqref{FGPA2}.
This diagram implies that at these values of $y$ and $\eps$ \fp{I} is IR repulsive point, \fp{II} is a saddle point, and \fp{III} is a spiral  attractor, in agreement with aforementioned analysis.

A general pattern of the stability of three fixed points in the
plane $(y,\varepsilon)$ is shown in Fig.~\ref{fig:Plain}.
The lines $y<0$, $\varepsilon=0$; $y=0$, $\varepsilon<0$; and $y=3\varepsilon/2$, $\varepsilon>0$
denote the boundaries of the domains, which
have neither gaps between them nor overlaps.
The crossover between two non-trivial fixed points {takes place} along the line $y=3\eps/2$, which is in accordance with~\cite{Ant04}.
The dotted line $y=9\varepsilon/5$ corresponds to $\lim_{\alpha\to\infty}R(\alpha,\eps)$~[see~\eqref{lim}] and 
indicates a boundary between areas in which the IR attractive point \fp{III} is a node ($3\varepsilon/2<y<9\varepsilon/5$) or a spiral ($y>9\varepsilon/5$) attractor at $\alpha=\infty$. 

The presence of the different IR attractive fixed points in the model~(\ref{eq:action})
implies that depending on the values $y$ and $\varepsilon$ the correlation functions of the model
in the IR region exhibit various types of scaling behavior.

 {The point \fp{II} [see~\eqref{gfp2}] is a fixed point of a
self-contained renormalizable local field theory, in which quadratic form~\eqref{quadlocal2} reduces to a single integral:}
 {
\begin{equation}
v'_iD_{ik}{v'}_k=g_{20}\nu_0^3\int \dRM t \!\int\! \dRM^d{\bm x}\, v^2(t,{\bm x}).
\end{equation}}
 {This regime corresponds to the ``compressible'' analog of model B in the
pioneering paper~\cite{FNS}. 
The authors interpret that this model describes a
macroscopic ``shaking'' of a fluid container(the idea suggested
by P.~C.~Martin, see footnote 15 in~\cite{FNS}), which is problem of a special (clear, practical) interest.
The unavoidable presence of a local term in \fp{III} [see~\eqref{gfp3}]
means that the non-local stirring force~\eqref{ff}, due to the
renormalization and the intristic non-linearity of the problem, gives rise to the
effective ``shaking'' effect.}

 {For the incompressible case, a new
scaling regime arises near the dimension $d=2$, see~\cite{HN96}. 
This regime formally corresponds to a fluid in
thermal equilibrium (model A in~\cite{FNS}). To avoid 
misunderstanding, it should be stressed that the true thermal noise does not come into play in
turbulence dynamics, but the non-local noise gives rise, due to the
non-linear nature of the whole problem, to effective thermal-like
noise.
The situation resembles ``effective turbulent diffusion,'' in which the
behavior of a particle in a turbulent environment resembles ordinary
diffusion, but with coefficients determined by the
charachteristics
of the turbulent flow.}

The corresponding critical dimensions $\Delta[F]\equiv\Delta_{F}$ for all
basic fields and parameters can be computed as series in 
a set of parameters $y$ and $\varepsilon$, where $y$ and $\varepsilon$ are assumed to be quantities of the same
order, i.e., $0<y/\eps<\infty$. 
 {If, for the real values $y=4$ and $\varepsilon=1$, the local point \fp{II} were IR attractive, the IR behavior of the full non-local model would be the same as for the local case described by the fixed point~\eqref{gfp2} with the dimensions given in~\eqref{Krit2}.  Our findings show that this is not the case, and the IR behavior is governed by the dimensions~\eqref{Krit3}}; see next subsection for details.

\subsection{IR attractive fixed points and critical dimensions}

In the leading order the IR asymptotic behavior of the (renormalized) Green functions $G^R$
satisfy the RG equation~\eqref{RG1} with the substitution
$g\to g_{*}$ for the full set of the
couplings $\left\{g_1,g_2,u,v\right\}$, see~\cite{Vasiliev,Amit}. This yields
\begin{equation}
\left\{
{\cal D}_{\mu} - \gamma_{\nu}^{*}{\cal D}_{\nu}- \gamma_{c}^{*}{\cal D}_{c}
+ \sum_{\Phi} N_{\Phi}\gamma_{\Phi}^{*}  \right\} \, G^{R} = 0.
\label{RGF}
\end{equation}
Here, $\gamma_{F}^{*}$ is the value of the anomalous dimension at the fixed point;
the summation over all types of the fields $\Phi$ is
implied. 
Equations of this type describe the
scaling with dilatation of the variables whose derivatives enter the
differential operator. 

From~\eqref{gammanu}~-- \eqref{gammag2}
one obtains that in the one-loop approximation the expressions for the anomalous dimensions $\gamma_{F}^{*}$ 
for the non-local point \fp{III} coincide with the case of $d=3$:
\begin{eqnarray}
\gamma_{\nu}^{*} &=& y/3, \quad
\gamma_{\phi}^{*}=-\gamma_{\phi'}^{*} =-y/6+  \mathcal{O}  (y^{2}),
\nonumber \\
\gamma_{c}^{*} &=& -y/12 + \mathcal{O}(y^{2}).
\label{Anom3}
\end{eqnarray}
The expression for $\gamma_{\nu}^{*}$ is exact due to the relation between $Z_\nu$ and $Z_{g_1}$, see~\eqref{eq:RGconst2}.
For the local point \fp{II} one obtains\footnote{There is a misprint in expression for $\gamma_{\nu}^{*}$ in published version.}
\begin{eqnarray}
\gamma_{\nu}^{*} &=& \varepsilon/2 + \mathcal{O}  (\varepsilon^{2}), \quad
\gamma_{\phi}^{*}=-\gamma_{\phi'}^{*} =-\varepsilon/4+  \mathcal{O}  (\varepsilon^{2}),
\nonumber \\
\gamma_{c}^{*} &=& -\varepsilon/8 +  \mathcal{O}  (\varepsilon^{2}).
\label{Anom2}
\end{eqnarray}

The canonical scale invariance is expressed by two relations
\begin{equation}
\left\{\sum _{\sigma}d_{\sigma}^k{\cal D}_{\sigma}-
d_{G}^k\right\}G^{R}=0 ,\quad
\left\{\sum _{\sigma}d_{\sigma}^{\omega }{\cal D}_{\sigma}-
d_{G}^{\omega }\right\}G^{R}=0 ,
\label{Canonic-Scl-Inv}
\end{equation}
where $\sigma$ is the full
set of all the arguments of $G^{R}$, $d^{k}$ and $d^{\omega}$ are 
canonical dimensions. 
In order to derive the scaling with fixed
``IR irrelevant'' parameters $\mu$ and $\nu$
one has to combine Eqs.~(\ref{RGF}) and~(\ref{Canonic-Scl-Inv})
such that the derivatives with respect to these
parameters are eliminated; see~\cite{turbo,Vasiliev}.
This  yields an equation of critical IR scaling for the model
\begin{eqnarray}
\left\{ - {\cal D}_{\bm x} + \Delta_{t} {\cal D}_{t}
+ \Delta_{c} {\cal D}_{c}+ \Delta_{m} {\cal D}_{m} -
\sum_{\Phi} N_{\Phi} \Delta_{\Phi} \right\} \, G^{R} = 0
\nonumber \\
\label{KS}
\end{eqnarray}
with the notation
\begin{equation}
\Delta_{F} = d^{k}_{F}+ \Delta_{\omega}d^{\omega}_{F} + \gamma_{F}^{*},
\quad \Delta_{\omega} = - \Delta_{t} =  2-\gamma_{\nu}^{*}.
\label{Krit}
\end{equation}
Here, $\Delta_{F}$ is the critical dimension of the quantity $F$, 
while $\Delta_{t}$ and $\Delta_{\omega}$ are the critical
dimensions of the time and the frequency. 

From Table~\ref{table1} and Eqs.~(\ref{Anom3}) and~(\ref{Anom2}) 
we see that for the local point \fp{II} critical dimensions take the form
\begin{align}
\Delta_{v}&=1-\varepsilon/2, &\Delta_{v'}&= d- \Delta_{v}, \nonumber\\
\Delta_{\omega}&=2-\varepsilon/2, &\Delta_{m}&=1,\nonumber\\
\Delta_{\phi}&=d-\Delta_{\phi'}=2-5\varepsilon/4, 
&\Delta_{c}&= 1- 5\varepsilon/8,
\label{Krit2}
\end{align}
whereas for the point \fp{III} they coincide with the case $d=3$, namely, 
\begin{align}
\Delta_{v} &=1-y/3,  &\Delta_{v'}&= d- \Delta_{v}, \nonumber\\
\Delta_{\omega}&=2-y/3, &\Delta_{m}&=1, \nonumber\\
\Delta_{\phi}&=d-\Delta_{\phi'}=2-5y/6, 
&\Delta_{c}&= 1- 5y/12.
\label{Krit3}
\end{align}

Expressions~\eqref{Krit2} and~\eqref{Krit3} implies that depending on the values $y$ and $\varepsilon$ correlation functions can exhibit different types of scaling behavior in the IR region (local regime \fp{II} or non-local regime \fp{III}) with different anomalous and critical dimensions.

\section{Advection of passive scalar fields} \label{sec:Ops1}

The analysis of the passive advection bears a close resemblance to   the case $d=3$ (see~\cite{AK14}), so we will restrict ourselves
to the main points.

\subsection{Field theoretic formulation of the model}
\label{sec:scheme}

Consider a passive advection of a scalar density field $\theta(x)\equiv \theta(t,{\bm x})$ {(e.g., density of a pollutant)}, which 
satisfies the stochastic differential equation
\begin{equation}
\partial _t\theta+ \partial_{i}(v_{i}\theta)=\kappa _0 \partial^{2} \theta + f_\theta.
\label{density1}
\end{equation}
{Another related problem}, which corresponds to the transformation $\partial_{i}(v_{i}\theta)\to(v_{i}\partial_{i})\theta$
in the left hand side, has an interpretation {of passive advection} of a tracer field (e.g., temperature, concentration of the
impurity particles, etc.); see~\cite{LL}. 
As usual, $\partial _t \equiv \partial /\partial t$,
$\partial _i \equiv \partial /\partial x_{i}$; 
$\kappa_0$ is the molecular
diffusivity coefficient, $\partial^{2}=\partial _i\partial _i$
is the Laplace operator, $ {v_i(x)}$ is the velocity field, which obeys Eq.~\eqref{NS},
and $f_\theta\equiv f_\theta(x)$ is a Gaussian noise with zero mean and given covariance
\begin{equation}
\langle  f_\theta (x)f_\theta (x')  \rangle = \delta(t-t')\, C({\bm r}/ {\cal L_\theta}), \quad
{\bm r}= {\bm x} - {\bm x}'.
\label{noise}
\end{equation}
The function $C({\bm r}/ {\cal L_\theta})$ in Eq.~\eqref{noise} is finite at $({\bm r}/ {\cal L_\theta})\to 0$ and rapidly
decays when $({\bm r}/ {\cal L_\theta})\to\infty$. Expression~\eqref{noise} brings about another large (integral) scale $ {\cal L_\theta}$,
related to the noise variable $f_\theta$, but henceforth we will not distinguish it {from}
its analog $ {\cal L}=m^{-1}$ in the correlation function  of the stirring force~(\ref{ff}). 
The noise is needed to maintain the steady state of the system, and in this respect it  accounts for the effects of initial and/or boundary conditions.

In the absence of the noise, Eq.~(\ref{density1}) acquires the form of a
continuity equation (conservation law); $\theta$ being the density of a
corresponding conserved quantity.  If the function in~(\ref{noise}) is chosen 
in such a way that its Fourier transform $C({\bm k})$
vanishes at ${\bm k}=0$, the fields $\theta$ or $\theta'$
remain to be conserved in statistical sense in the presence of the
external stirring.

The advection of {scalar} fields in the case of Kraichnan's rapid-change velocity ensemble
were thoroughly studied~\cite{VM,Avell,tracer2,tracer3,tracer,RG1,AH,AG};
the case of Gaussian velocity statistics with finite correlation time
was studied in~\cite{AKens,AK2}.

If velocity  {$v_i$} obeys the stochastic Navier-Stokes equation~\eqref{NS}, the problem~(\ref{density1}),~(\ref{noise}) is tantamount
to the field theoretic model of the full set of fields
$ {\widetilde{\Phi}}\equiv\{\theta', \theta,  {v_i', v_i}, \phi', \phi  \}$
and the action functional
\begin{equation}
{\cal S}( {\widetilde{\Phi}})= {\cal S}_{\theta}(\theta', \theta,  {v_i}) +
{\cal S}_{ {{\bm v}}}( {v_i', v_i}, \phi', \phi).
\label{Fact}
\end{equation}
The advection-diffusion component
\begin{eqnarray}
{\cal S}_{\theta} (\theta', \theta,  {v_i}) = \frac{1}{2} \theta' D_{f} \theta' +
\theta' \left[ - \partial_{t}\theta -
\partial_{i}(v_{i}\theta) +\kappa _0 \partial^{2} \theta \right]
\nonumber \\
\label{Dact}
\end{eqnarray}
in Eq.~\eqref{Fact} is the De Dominicis-Janssen action for the stochastic problem
(\ref{density1}), (\ref{noise}) at fixed $ {v_i}$, while the second term is given by
(\ref{eq:action}) and represents the velocity statistics; $D_{f}$ is the
correlation function ~(\ref{noise}), all the required
integrations and summations over the vector indices are assumed.

\begin{figure}[t]
   \centering
   \begin{tabular}{c}
     \includegraphics[width=6.1cm]{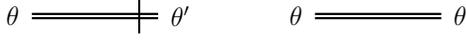}
      \end{tabular}
   \caption{Graphical representation of the bare propagators $\langle \theta \theta' \rangle _0$ and $\langle \theta \theta \rangle _0$.}
   \label{P2}
\end{figure}

\begin{figure}[t]
  \includegraphics[width=4.8cm]{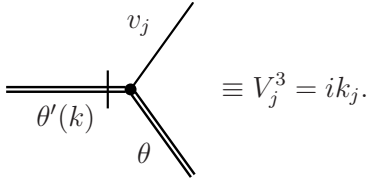}
  \caption{Graphical representation of the interaction vertex $V^3_j$.}
  \label{V3}
\end{figure}

In addition to~(\ref{lines}), the diagrammatic technique in the full
problem involves a new vertex $V^3_j=-\theta'\partial_{j}(v_{j}\theta)$ 
and two new propagators
\begin{eqnarray}
\langle \theta \theta' \rangle _0=  \overline{\langle \theta' \theta \rangle}_0 &=&
\frac{1} {-{\rm i}\omega +\kappa _0 k^2},
\nonumber \\
\langle \theta \theta \rangle _0 &=& \frac {C({\bm k})}
{\omega^{2} +\kappa_0^{2} k^4}.
\label{lines3}
\end{eqnarray}
From now  a double solid line without a slash
denotes the field $ \theta $, and a
double solid line with a slash
corresponds {to} the field $ \theta'$; see Fig.~\ref{P2}.
The vertex $V^3_j$ is depicted on Fig.~\ref{V3} and in the momentum representation
is given by
\begin{eqnarray}
{V^3_{j} ({\bm k})} = ik_{j},
\label{vertex1}
\end{eqnarray}
where ${\bm k}$ is the momentum carried by the field $\theta'$.

\subsection{Renormalization of the model}

Canonical dimensions of the new fields and parameters of the full 
model~(\ref{Fact}) can be found in Table~\ref{table1}.
The formal index of UV divergence (\ref{index}) remains valid, but now
the  summation has to run over the full set of six fields
$ {\widetilde{\Phi}}\equiv\{\theta', \theta,  {v_i', v_i}, \phi', \phi  \}$. Rules (1)~-- (6)
from Sec.~\ref{sec:Canon} have to be generalized and augmented:
\begin{enumerate}[(1)]
\setcounter{enumi}{6}
 \item All the {1-irreducible} Green functions without any response fields
 {$\widetilde{\Phi}'$} vanish identically and require no counterterms.
 \item
 Using the integration by parts the derivative at the vertex
$-\theta'\partial_{i}(v_{i}\theta)$ can be moved onto the field $\theta'$,
therefore, Eq.~\eqref{real} is modified:
\begin{equation}
\delta_{\Gamma}' = \delta_{\Gamma}- N_{\phi} - N_{\theta'}.
\label{realD}
\end{equation}
Since the field $\theta'$ 
can enter the counterterms only in the form of spatial derivatives,
the counterterm $\theta'\partial_{t}\theta$ to the {1-irreducible}
Green function $\langle\theta'\theta\rangle$ with
$\delta_{\Gamma}=2$, $\delta_{\Gamma}'=1$ is forbidden.
Also this requires that the counterterms to the
 {1-irreducible} function  {$\langle\theta'v_i\theta\rangle$} with
$\delta_{\Gamma}=1$, $\delta_{\Gamma}'=0$ necessarily reduce to the
form $\theta'\partial_{i}(v_{i}\theta)$.
Galilean symmetry allows them to enter the
counterterms only in the form of invariant combination
$\theta'\nabla_{t}\theta$. Hence, they are also forbidden.

\item As a consequence of the linearity of the original stochastic equation~(\ref{density1}) with respect to the field $\theta$ 
one obtains that for any {1-irreducible} function the relation $N_{\theta'}- N_{\theta}=2N_{0}$ is valid
(here $N_{0}\geq0$ is the total number of bare propagators
$\langle\theta\theta\rangle_0$ entering the diagram). 
This fact is very important for 
the renormalizability of the model: 
without the restriction $N_{\theta}\le N_{\theta'}$, 
 the infinite number of superficially divergent {1-irreducible} functions
$\langle \theta'\theta\dots\theta\rangle$ would proliferate, and hence the lack of
renormalizability would follow.
\end{enumerate}

From these rules we finally conclude that
superficial divergences can be present only in the {1-irreducible}
Green function $\langle\theta'\theta\rangle$ with the only
counterterm
$\theta'\partial^{2}\theta$. It is naturally reproduced as multiplicative
renormalization of the diffusion coefficient, $\kappa_{0}=\kappa Z_{\kappa}$.
No renormalization of the fields $\theta'$ and $\theta$ is needed:
$ Z_{\theta'}=Z_{\theta}=1$. Altogether, the renormalized analog of the action
functional~(\ref{Fact}) takes the form

\begin{equation}
{\cal S}^{R}( {\widetilde{\Phi}})= {\cal S}_{\theta}^{R}(\theta', \theta,  {v_i}) +
{\cal S}^{R}_{ {\bm v} }( {v_i', v_i}, \phi', \phi),
\label{FactR}
\end{equation}
where ${\cal S}^{R}_{\bm v}$ is the action~(\ref{Raction}), 

\begin{equation}
{\cal S}^{R}_{\theta} (\theta', \theta,  {v_i}) = \frac{1}{2} \theta' D_{f}\theta' 
+ \theta' \left[ - \partial_{t}\theta -
\partial_{i}(v_{i}\theta) +\kappa Z_{\kappa} \partial^{2} \theta \right];
\label{DactR}
\end{equation}
$D_f$ here stands for the covariance of the stochastic force given by the Eq.~\eqref{noise}.


\subsection{Calculation of the diagram, fixed points and critical dimensions}
\label{sec:AmonDim}

The one-loop approximation for the 1-irreducible response function $\langle\theta'\theta\rangle$ can be formally written as
\begin{align}
   \Gamma_{\theta'\theta} & = + i\omega - \kappa_0 p^{2} +  \raisebox{-1ex}{ \includegraphics[width=2.6truecm]{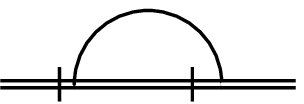}},
 \label{Dyson}
\end{align}
where, as earlier in the expressions~\eqref{eq:vsv}~-- \eqref{eq:vsvs}, ${\bm p}$ stands for an external 
momentum entering the diagram; 
the single solid line denotes the bare propagator $\langle vv\rangle_0$ from~(\ref{lines}), the double solid line with a slash
denotes the bare propagator $\langle \theta \theta' \rangle _0$ from~(\ref{lines3}), the slashed end corresponds to the field $\theta'$.
The interaction vertex with three attached fields $\theta'$, $\theta$ and $v$ contains the
factor~(\ref{vertex1}).
 
The renormalization constant $Z_\kappa$ should be chosen as
\begin{align}
Z_{\kappa} = 1 - \frac{1}{2dw}
\left[ \frac{d-1}{w+1} + \frac{\alpha(u-w)}{u(u+w)^{2}}
\right]\frac{g_1}{y} \nonumber \\
- \frac{1}{2dw}
\left[ \frac{d-1}{w+1} + \frac{u-w}{u(u+w)^{2}}
\right]\frac{g_2}{\varepsilon},
\label{Zk}
\end{align}
where we introduced the new dimensionless coefficient $w_0=\nu_0/\kappa_0$ with $\nu_{0}$ from~(\ref{NS}) and its renormalized analog $w$.
The corresponding anomalous dimension is
\begin{align}
\gamma_{\kappa} =   \frac{1}{2dw}
\left[ \frac{d-1}{w+1} + \frac{\alpha(u-w)}{u(u+w)^{2}}
\right]g_1 \nonumber \\ 
+ \frac{1}{2dw}
\left[ \frac{d-1}{w+1} + \frac{u-w}{u(u+w)^{2}}
\right]g_2,
\label{gk}
\end{align}
with the possible corrections coming from higher orders terms;  {see Appendix~\ref{app2a} for details.}

The function $\beta_{w} = \widetilde {\cal D}_{\mu} w$ for the new
parameter $w$ takes the form
\begin{equation}
\beta_{w} = - w\gamma_{w} = w (\gamma_{\nu}-\gamma_{\kappa}),
\label{betaw}
\end{equation}
see Sec.~\ref{sec:rgrg}. Now the 
coordinates $\left\{g^*\right\}$ of the fixed points \fp{II} and \fp{III} [see~\eqref{gfp2} and~\eqref{gfp3}]
are substituted into the equation $\beta_{w} =0$ at $d=4$. We can rewrite the expression for $\gamma_{\nu}-\gamma_{\kappa}$ at $u=1$:
\begin{align}
\gamma_{\nu}-\gamma_{\kappa}|_{u=1}&=\frac{w-1}{16w(w+1)^2} \left[g_1(3w^2+9w+2\alpha+6)\right.\nonumber \\
&\left.+g_2(3w^2+9w+8)\right].
\label{u1}
\end{align}
From Eq.~\eqref{u1} it is clear that the only positive solution for both \fp{II} and \fp{III} is
\begin{equation}
w^* =1.
\label{wfp}
\end{equation}

The functions~(\ref{betagw}) do not depend on $w$. Therefore, a new eigenvalue of the matrix~(\ref{Omega}), corresponding to this parameter, coincides
with the diagonal element $\partial\beta_{w}/\partial w$ at the point $\left\{g\right\}=\left\{g^*\right\}$:
\begin{align}
\lambda_w&=\frac{5\varepsilon}{6}>0 \quad\text{for \fp{II}}; \nonumber \\
\lambda_w&=\frac{2y}{3}+\frac{4\alpha y(y-\varepsilon)}{3[y(\alpha+2)-3\varepsilon]}>0 \quad\text{for \fp{III}}.
\label{wir}
\end{align}

From the inequalities~\eqref{wir} it follows that the fixed points with the coordinates~\eqref{gfp2} and~\eqref{gfp3} 
and $w_{*}=1$ are IR attractive
in the full space of couplings $\left\{g_1,g_2,u,v,w\right\}$ and govern the IR asymptotic
behavior of the full-scale model~(\ref{Fact}).

The critical dimensions of the passive fields $\theta$ and $\theta'$ are obtained from
Table~\ref{table1} and Eq.~(\ref{Krit}) for $\Delta_{\omega}$. 
For \fp{II} they are 
\begin{equation}
\Delta_{\theta}= -1+\varepsilon/4, \quad \Delta_{\theta'}= d+1 -\varepsilon/4;
\end{equation}
for \fp{III} they are the same as in the case $d=3$, namely,
\begin{equation}
\Delta_{\theta}= -1+y/6, \quad \Delta_{\theta'}= d+1 -y/6.
\end{equation}

\subsection{Renormalization and critical dimensions of composite operators} \label{sec:Ops}

In the following, the central {role} is played by composite fields
(``composite operators'') built solely from the basic fields $\theta$: 
\begin{equation}
\label{comp}
F(x)=\theta^{n}(x).
\end{equation}
In general, a local composite
operator is a polynomial constructed from the primary fields
$\Phi(x)$ and their finite-order derivatives at a single space-time point
$x=(t,{\bm x})$. Due to a coincidence of the field arguments, new UV
divergences arise in the Green functions with such objects~\cite{Zinn,Vasiliev}.

The total canonical dimension of an arbitrary 1-irreducible Green function
$\Gamma = \langle F\,\Phi\dots\Phi\rangle$
that includes one composite operator $F$ and arbitrary number of
primary fields $\Phi$ (the formal index of UV divergence) is given by the relation
\begin{equation}
\label{D-Comp}
d_{\Gamma}=d_F-\sum_\Phi N_\Phi d_\Phi,
\end{equation}
where $N_{\Phi}$ is the number of the field $\Phi$ {entering  $\Gamma$},
$d_{\Phi}$ is the total canonical dimension of the given field $\Phi$, $d_{F}$ is the canonical
dimension of the operator.

In the process of renormalization operators can mix with each other,
\begin{equation}
F_{i}=\sum _{j} Z_{ij} F_{j}^{R},
\end{equation}
and $Z_{ij}$ is the renormalization matrix. But in the simplest case of the operators~\eqref{comp} the matrix $Z_{ij}$ is diagonal,
i.e., $F(x)= Z_{F} F^{R}(x)$. In particular, this means that
the critical dimension of the operator is
given by the expression~(\ref{Krit}).

Superficial UV divergences, whose removal requires counterterms, can be
present only in those functions $\Gamma$ for which the index of
divergence $d_{\Gamma_{N_\Phi}}$ is a non-negative integer.
For the operators of the form~\eqref{comp} one has $d_F=-n$.
Due to the linearity of our model in $\theta$,
the number of fields $\theta$ in any 1-irreducible function with the operator
$F(x)$ cannot exceed
their number in the operator itself.
Thus, from the analysis of Eq.~(\ref{D-Comp}) it follows that
the superficial divergence can only be present in the
1-irreducible function with $N_{\theta}=n$ and $N_{\Phi}=0$ for all other types of the fields
$\Phi$. 
For this function $\delta_{\Gamma}=0$ and the corresponding counterterm takes the form
$\theta^{n}(x)$; hence, the operators in~\eqref{comp} are multiplicatively renormalizable,
$F(x) = Z_{n} F^{R} (x)$.

Let us {introduce} $\Gamma_n(x;\theta)$: the $\theta^{n}$ term of the
expansion in $\theta(x)$ of 
the generating functional of the
 {1-irreducible} Green functions with one composite operator $F(x)$
and any number of fields $\theta$:
\begin{align}
\Gamma_{n}(x;\theta) =  &\int \dRM x_{1} \cdots \int \dRM x_{n}
\langle F(x) \theta(x_{1})\cdots\theta(x_{n})\rangle \nonumber \\
&\times \theta(x_{1})\cdots\theta(x_{n}).
\label{Gamma1}
\end{align}
The renormalization constants $Z_n$ are determined by the requirement that the {1-irreducible} functions~\eqref{Gamma1}
be UV finite in the renormalized theory.

The one-loop approximation for the 1-irreducible function $\Gamma_{n}(x;\theta)$ can be formally written as
\begin{align}
\Gamma_{n}(x;\theta)= F(x) + \frac{1}{2}  \vcenter{\hbox
{\includegraphics [width=.09\textwidth,clip]{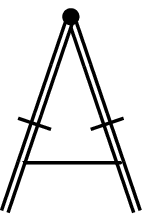}}}.
\label{Gamma2}
\end{align}
The first term in Eq.~\eqref{Gamma2} is the tree (loop-less) approximation, the 
double solid lines with a slash
denotes the propagators $\langle \theta \theta' \rangle$, the
single solid line corresponds to the propagator $\langle vv \rangle$,
$1/2$ is the symmetry coefficient of the given graph,
and the dot with two attached lines in the top of the diagram denotes the operator vertex, i.e.,
the variational derivative
\begin{align}
V(x;x_{1},x_{2})&=\delta^{2}F(x)/{\delta\theta(x_{1})\delta\theta(x_{2})} \nonumber \\
&=n(n-1)\, \theta^{n-2}(x)\, \delta(x-x_{1})\delta(x-x_{2}).
\label{Vae}
\end{align}
A contribution of a specific diagram into the functional~\eqref{Gamma2}
for any composite operator $F$ is represented in
the form
\begin{equation}
\label{Diag-General}
\Gamma_{n} = V\times I\times
\theta \dots \theta,
\end{equation}
where $V$ is the vertex factor given by Eq.~\eqref{Vae},
$I$ is the diagram itself, and the product $\theta \dots \theta$ corresponds
to the external tails.

The renormalization constants $Z_{n}$ are found from the requirement that the
renormalized analog $\Gamma_{n}^{R}=Z_{n}^{-1}\Gamma_{n}$ of the function
(\ref{Gamma1}) be UV finite in terms of renormalized parameters and take a form
\begin{equation}
Z_{n} = 1 + \frac{n(n-1)}{4wu(u+w)}\left(\frac{\alpha g_1}{y}+\frac{g_2}{\varepsilon} \right);
\label{Zn}
\end{equation}
see Appendix~\ref{app2b} for details.
The corresponding anomalous dimensions are 
\begin{equation}
\gamma_{n} = - \frac{n(n-1)}{4wu(u+w)}\left(\alpha g_1+g_2\right),
\label{gaman}
\end{equation}
with higher-order corrections in $g_1$ and $g_2$.

The critical dimensions of the operators $\theta^{n}$ from the
expression~(\ref{Krit}) are readily derived
\begin{equation}
\Delta[\theta^{n}]= n\Delta_{\theta} + \gamma_{n}^{*}.
\label{KrOp}
\end{equation}
Substituting the fixed-point values \fp{II} and \fp{III} into Eq.~(\ref{gaman}) finally
gives the critical dimensions
\begin{align}
\Delta\left[\theta^n\right]&=-n+\frac{n\varepsilon}{4}-\frac{n(n-1)}{3}\varepsilon
\label{KrOp1}
\end{align}
for the point \fp{II};
\begin{align}
\label{KrOp2}
\Delta\left[\theta^n\right]&=-n+\frac{ny}{6}-\frac{2n(n-1)}{3}\frac{\alpha y(y-\varepsilon)}{y(\alpha+2)-3\varepsilon}
\end{align}
for the point \fp{III}.
Both the expressions~\eqref{KrOp1} and~\eqref{KrOp2} assume higher-order corrections in $y$ and $\varepsilon$.
For both cases, \fp{II} and \fp{III}, the dimensions are negative, i.e., ``dangerous'' in the 
sense of operator product expansion~\cite{Vasiliev,turbo},
and decrease as $n$ grows. 

The latter result for \fp{III} is in agreement with previously known result~\cite{AK14} for the analysis near three-dimensional space $d=3$:
\begin{align}
\Delta\left[\theta^n\right]&=-n+\frac{ny}{6}-\frac{n(n-1)}{6}\frac{\alpha dy}{(d-1)},
\end{align}
which at $d=4$ reads
\begin{align}
\label{KrOp2-1}
\big.\Delta\left[\theta^n\right]\biggl|_{d=4}&=-n+\frac{ny}{6}-\frac{2\alpha y}{9}n(n-1).
\end{align}
Expanding the  expression~\eqref{KrOp2} in $y$ at fixed (not small) value  $\eps=1$ (which corresponds to $d=3$) gives
\begin{align}
\label{KrOp2-2}
\Delta\left[\theta^n\right]&=-n+\frac{ny}{6}-\frac{2\alpha y}{9}n(n-1)+\mathcal{O}(y^2).
\end{align}
From the expressions~\eqref{KrOp2-1} and~\eqref{KrOp2-2} it follows that the expression~\eqref{KrOp2}, obtained as a result of the double $y$ and
$\varepsilon$ expansion near $d=4$, may be considered as a certain partial infinite resummation of the ordinary $y$ expansion. This
resummation significantly improves the situation at large $\alpha$~-- now we do not have the pathology when the critical dimensions $\Delta\left[\theta^n\right]$ grow with $\alpha$ without a bound 
and also that the fixed point ceases to exist at the single value $\alpha=\infty$.

\subsection{Operator product expansion and anomalous scaling}
\label{sec:OPE}

The measurable quantities and, therefore, the objects of interest are 
equal-time pair correlation functions of two (UV finite) 
renormalized local composite operators
$F_{1,2}(x)$. 
From the {dimensional} considerations (see Table~\ref{table1}) it follows that
\begin{equation}
\langle F_{1}(t,{\bm x}_{1}) F_{2}(t,{\bm x}_{2}) \rangle =
\nu^{d^{\omega}_{F}} \mu^{d_{F}} f(\mu r, mr, c/\mu\nu),
\nonumber \\  {}
\label{Parr0}
\end{equation}
where $d^{\omega}_{F}$ and $d_{F}$ are the frequency and total canonical dimensions of the
correlation function, $r=|{\bm x}_{2}-{\bm x}_{1}|$,
and $f$ is a function of  dimensionless
variables. 

If the correlation function~(\ref{Parr0}) is multiplicatively renormalizable, in the IR region it fulfills
the differential equation~\eqref{KS}, which 
describes the IR scaling behavior. That is, the behavior of the function $f$ for $\mu r \gg1$ 
is determined by the IR attractive fixed points \fp{II} and \fp{III} of the
RG equation. 
A solution of this equation leads to the following asymptotic expression:
\begin{equation}
\langle F_{1}(t,{\bm x}_{1}) F_{2}(t,{\bm x}_{2}) \rangle \simeq
\nu^{d^{\omega}_{F}} \mu^{d_{F}}(\mu r)^{-\Delta_{F}}
h[mr, \bar{c}(r)].
\label{Parr}
\end{equation}
Here, $\Delta_{F}$ is the critical dimension of the correlation function,
given by a simple sum of the dimensions of the operators; 
$h$ is an unknown scaling function with completely (both canonically and critically) dimensionless arguments, and
$\bar{c}(r)$ is invariant speed of sound.

For the composite operator $F(x)=\theta^n(x)$, Eq.~(\ref{Parr}) yields
\begin{equation}
\langle \theta^{p}(t,{\bm x}_{1}) \theta^{k}(t,{\bm x}_{2}) \rangle \simeq
\mu^{-(p+k)} (\mu r)^{-\Delta_{p}-\Delta_{k}} h_{pk}[mr,\bar{c}(r)],
\label{Parr2}
\end{equation}
where the critical dimensions $\Delta_{n}$ for two scaling regimes are given by Eqs.~(\ref{KrOp1}) and~(\ref{KrOp2}). 

The representation~\eqref{Parr2} holds for $\mu r\gg1$ and any fixed value of $mr$.
The inertial-convective range
$l\ll r\ll  {\cal L}$  corresponds to the additional condition $mr\ll1$. 
Behavior of the function $h$ at $mr\to0$ can be
studied by means of the operator product expansion; see~\cite{Zinn,Vasiliev}.
According to the OPE, the equal-time product $F_{1}(x_1)F_{2}(x_2)$
of two renormalized operators for
${\bm x}\equiv ({\bm x_1} + {\bm x_2} )/2 = {\rm const}$ and
${\bm r}\equiv {\bm x_1} - {\bm x_2}\to 0$ takes the form
\begin{equation}
F_{1}(t,{\bm x}_{1}) F_{2}(t,{\bm x}_{2}) \simeq \sum_{F} C_{F}[mr, \bar{c}(r)]
F(t,{\bm x}),
\label{OPE}
\end{equation}
where $C_{F}$ are numerical coefficient
functions analytical in $mr$ and $\bar{c}(r)$ and 
$F$ are all possible renormalized local composite operators allowed by the symmetry.

The correlation function~(\ref{Parr}) 
is obtained by averaging~(\ref{OPE}) with the weight
$\exp \S_{R}$, where $\S_{R}$ is the renormalized action functional~(\ref{Fact}). 
Mean values $\langle F(x)\rangle~\propto~(mr)^{\Delta_{F}}$
appear in the right hand side.  Their asymptotic behavior
at small $m$ is found from the corresponding RG equations and
takes the form

\begin{equation}
\langle F(x) \rangle \simeq m^{\Delta_{F}} q[\bar{c}(1/m)],
\label{Mean}
\end{equation}
with another set of scaling functions $q$. Since the diagrams of the perturbation theory have finite
limits both for $c\to\infty$ and $c\to 0$, we may assume that the functions
$q(c)$ are restricted for all
values of $c$ and can be estimated by some constants. 
Moreover, for the invariant variable $\bar{c}(r)$ IR asymptotic behavior together with requirement of its dimensionless gives 
\begin{equation}
{\bar{c}(r) }= c (\mu r)^{\Delta_{c}} / (\mu\nu),
\label{rms}
\end{equation}
where $c$ is renormalized speed {of} sound.
Thus, $\bar{c}(1/m) \sim c m^{-\Delta_{c}}$. Taking into account~\eqref{Krit3}, for the non-local scaling regime \fp{III}
one obtains that for $y>12/5$ (i.e., including the most realistic case $y\to4$) the argument $cm^{-\Delta_{c}}$ becomes small for fixed $c$ and
$m\to0$, and the function $q$ can be replaced by its finite limit value
$q(0)$. For the local scaling regime \fp{II} from~\eqref{Krit2} it follows that {as} $\varepsilon\to 1$
 the function $q$ can be replaced by its finite limit value
$q(\infty)$.
From these two remarks we conclude that in the IR range for both the local and non-local scaling regimes up to a different constants
we can write
\begin{equation}
\langle F(x) \rangle \sim m^{\Delta_{F}}.
\label{Mean2}
\end{equation}
Combining the RG representation~(\ref{Parr2}) with the information gained from the
OPE~(\ref{OPE}) and Eq.~(\ref{Mean2})
gives the desired asymptotic  behavior of the scaling functions
\begin{equation}
h[mr, c(r)] \simeq \sum_{F} A_{F} [mr, c(r)] \, (mr)^{\Delta_{F}},
\label{Fin}
\end{equation}
where the summation runs over all the Galilean invariant scalar operators (including operators with derivatives, etc.), with
the coefficient functions $A_{F}$ analytical in their arguments. The leading contribution {in} the sum~\eqref{Fin} is given 
by the operator with the lowest (minimal) critical dimension; others can be considered as corrections. 
The anomalous scaling (i.e., singular behavior as $mr\to0$) results from the
contributions of the operators with negative critical dimensions.
From~\eqref{KrOp1} and~\eqref{KrOp2} it is easily seen that for both scaling regimes all the operators $\theta^{n}$ have negative dimensions,
and the spectrum of their dimensions is not restricted
from below.

Fortunately, due to the linearity of the initial stochastic equation~\eqref{density1}
in the field $\theta$, the number of such fields in the right hand side of the expression~\eqref{OPE}
cannot exceed their number in the left hand side. Thus, 
for a given correlation function
only a finite number of those operators
can contribute to the OPE. For  the correlation functions~(\ref{Parr2}) these operators are those for which
$n\le p+k$. The leading term of the behavior as $mr\to0$ 
is given by the operator with the maximum possible $n=p+k$ and without any derivatives, so 
the final expression takes the form
\begin{equation}
\langle \theta^{p}(t,{\bm x}_{1}) \theta^{k}(t,{\bm x}_{2}) \rangle \simeq
\mu^{-(p+k)} (\mu r)^{-\Delta_{p}-\Delta_{k}} (mr)^{\Delta_{p+k}}.
\label{FinF}
\end{equation}

The fact that the leading term in the OPE is given by the operator from
the same family with the summed exponent together with 
inequality $\Delta_{p}+\Delta_{k}>\Delta_{p+k}$ can be interpreted as the statement
that the correlations of the scalar field in the model~\eqref{density1} show
multi-fractal behavior; see~\cite{DL}.

\section{Conclusion} \label{sec:Conc}

In this paper, which is an extension of~\cite{ANU97,AK14}, the stochastic Navier-Stokes equation for a compressible fluid 
was studied using the field theoretic approach.
In contrast to previous studies, we analyzed the model  
near the special space dimension $d=4$, where the model possesses an additional 
UV divergence in the 1-irreducible Green function  {$\left\langle v_i'v_j'\right\rangle$}. 
This feature significantly affects 
both technical aspects and results of the RG analysis. 
In particular, it necessitates the renormalization group technique with a double expansion scheme.
In the one-loop approximation, the model possesses two attractive non-trivial
fixed points in the IR region, i.e., two possible non-trivial scaling regimes~-- a local one, denoted \fp{II} in 
the text, and a non-local one, \fp{III}. 
These points depend on the exponent $y$ and on 
$\eps=4-d$, the deviation of the space dimension $d$ from its special value $4$.

Analysis at $d=3$, which finds only one non-trivial fixed point corresponding to the non-local scaling regime~\cite{ANU97,AK14} should therefore be regarded as incomplete.
The crossover between the local and non-local regimes occurs along the line $y=3\varepsilon/2$, which is in accordance with~\cite{Ant04}.
The new (local) regime, which arises at $d = 4$, continuously  moves to $d = 3$ as $\varepsilon~\to~1$.
Nevertheless, the quantitative RG analysis, based on the one-loop approximation, shows that for the real values of the parameters 
$y=4$ and $\varepsilon=1$ the new local point \fp{II} is not IR attractive, but the non-local point \fp{III} is. This
finding confirms the RG analysis in~\cite{ANU97,AK14}, done within the single expansion in $y$. 
 {However, the situation may change at the two-loop level, where, for example, the 
areas of stability of two different fixed points may overlap and the 
choice of fixed point, which defines asymptotic behavior, will depend on the initial data $g_{10}, g_{20}, u_0, v_0$, and $w_0$.
Herewith, the local point \fp{II} describes the system near thermal equilibrium and is valid (IR attractive) for all $y$ and $\varepsilon$ if the pumping of energy by large-scale eddies 
is absent, i.e., if $g_{10}=0$.}

We also analyzed the model of passive scalar advection of density field by this velocity ensemble.
The full stochastic problem can be formulated as a field theoretic model,
which is multiplicatively renormalizable. The new parameter $\kappa$ does not affect the RG functions of the Navier-Stokes equation itself, 
so the critical behavior of this model is also described by two fixed points, a local one and non-local one.
The inertial range ($l\ll r\ll  {\cal L}$) behavior of correlation functions 
was studied using the OPE technique. The existence
of anomalous scaling, i.e., singular power-like dependence on the integral scale~$ {\cal L}$, was established. 
The corresponding anomalous exponents were identified
with critical dimensions of certain composite operators
and calculated in the leading one-loop approximation. 

The results of this study are especially significant at large values of $\alpha$ (purely potential random force).
In contrast to analysis near $d=3$, in the present case the anomalous dimensions of the composite operators~\eqref{KrOp1} and~\eqref{KrOp2} do not 
grow with $\alpha$ without a bound. This is a consequence of eliminating the poles in $\varepsilon$ near $d=4$, which leads 
to a significant improvement of calculated expressions for critical dimensions near physical value $d=3$.
A previous study~\cite{AK14} suggested that the real expansion parameter is $\alpha y$ rather than $y$, therefore, any 
finite order of this ($\alpha y$) expansion is not suitable for studying the behavior at large $\alpha$. According to this
observation, it is necessary to perform a resummation assuming that $y$ is small and $\alpha y\sim 1$. 
Expression~\eqref{KrOp2} obtained in this study provides an example of such resummation. 
It works well at large $\alpha$ being not expanded in $y$, and the first 
term of this expansion coincides with the answer presented in~\cite{AK14}; see the expressions~\eqref{KrOp2-1}~-- \eqref{KrOp2-2}.
The hypothesis that the scaling regimes undergo a qualitative changeover, possibly accompanied
by phase transition to a purely chaotic state, was presented in previous study~\cite{ANU97}
based on the observation that for some large value of $\alpha$ the points \fp{II}
and \fp{III} disappear or lose their stability. From the expression~\eqref{KrOp2} it follows that this hypotesis is not confirmed. 
The consideration of the present model near $d=4$
is similar to the RG analysis of the Navier-Stokes equation for incompressible fluid near $d = 2$, where additional renormalization near the special space dimension $d=2$ improves the agreement of the predicted Kolmogorov constant  with experimental results~\cite{AHKV05}. 

Double $y$ and $\varepsilon$ expansion near $d=4$ provides 
an additional interesting opportunity~-- it allows to
analyze whether a non-local fixed
point is a node or a spiral attractor. 
The anomalous exponents do not depend on the type of attractor, but the behavior of the RG flow is interesting itself. Depending
on the values of the parameters $y$ and $\varepsilon$ the point \fp{III} might be a spiral 
attractor if $\alpha>\frac{1}{5} (-7 + 3 \sqrt{6})\approx0.07$. 
At their physically relevant values, i.e., $y=4$ and $\varepsilon=1$,
the point \fp{III} becomes a spiral attractor if $\alpha>\frac{5}{176} (-26 + 9 \sqrt{15})\approx0.25$.

It would be very interesting to go beyond the
one-loop approximation and to examine the existence, stability and
$\alpha$-dependence of fixed points at the two-loop
level, which seems to be a technically difficult task. 
In addition, it would be very interesting to investigate scalar admixture
in the case of a tracer field or passively advected vector fields.
Another very important task is to develop the compressible Navier-Stokes equation near $d=2$. Such analysis may reveal 
additional types of IR behavior or another dependence on parameters like $\alpha$, viscosity ratios, etc.
These studies are underway and are left for the future.

\section*{Acknowledgments}

The authors are indebted to Loran~Ts.~Adzhemyan, Michal Hnati\v{c}, Juha Honkonen, Mikhail V.~Kompaniets, Nikita M.~Lebedev, Mikhail Yu.~Nalimov, and Viktor \v{S}kult\'ety for discussions and Maria, John, and Aviva Bloom for reading the manuscript.

The work was supported by VEGA grant No.~ 1/0345/17 of the Ministry
of Education, Science, Research and Sport of the Slovak Republic,
by the Russian Foundation for Basic
Research within the Project 16-32-00086, 
and by the Ministry of Education and Science of Russian Federation (the Agreement number 02.a03.21.0008).
N.~M.~G. acknowledges the support from the Saint Petersburg Committee of Science and High School.
N.~M.~G. and M.~M.~K. were also supported by Dmitry Zimin's Dynasty Foundation.

\appendix

\section{Calculation of the diagrams for Navier-Stokes stochastic equation}
\label{app1}

This section contains detailed calculations of the diagrams, defining the renormalization constants $Z_1$~-- $Z_6$ (see Sec.~\ref{Zs}).
All calculations are performed in the analytical regularization and the MS scheme. All diagrams
are calculated in arbitrary space dimension $d$, and only poles in $y$ and $\eps=4-d$ are presented in the results. 
The renormalization constants obtained this way do not depend on the parameter $c_0 \sim c$, so that it is possible to set $c_0 = 0$ in the propagators [see~\eqref{lines}] in all the cases, in which some quantity is not proportional to $c_0$. If some quantity is proportional to $c_0$, we may set $c_0=0$ after we have obtained the needed power of it. This means that we may set $c_0=0$ in all denominators, preserving them in numerators. The situation is similar to {calculations} of 
critical exponents in models of critical behavior, which can be performed in the ``massless'' models: we may consider $c_0$ to play a similar role as 
$\tau\propto T-T_c$ in $\phi^4$ model. 
In the MS scheme, the renormalization constants do not depend on $\tau$ and can be calculated directly at the critical point $\tau = 0$; see~\cite{ANU97,Vasiliev}.

\vspace{-1.8em}
\begin{widetext}
\subsection{The diagram with $d_\Gamma=0$}
\label{sec:c}

Start with the simplest graph for which $d_\Gamma=0$ and which appears in Eq.~\eqref{eq:vsvs}:
\begin{eqnarray}
D_1=\raisebox{-1ex}{ \includegraphics[width=2.5truecm]{VsVs.eps}}.
\end{eqnarray}
The corresponding analytical
expression reads
\begin{eqnarray}
\label{Expr-D1}
D_1&=&(-i)^2\int\frac{\dRM\omega}{2\pi}\int\frac{\dRM^d{\bm k}}{(2\pi)^d} {(k_b\delta_{ia}-k_a\delta_{bi})(k_c\delta_{jd}-k_d\delta_{jc})} \nonumber \\
&\times&\left[P_{ac}({\bm k})\frac{g_{10}\nu_0^3k^{4-d-y}+g_{20}\nu_0^3}{|\epsilon_1(k)|^2}+
Q_{ac}({\bm k})\left(\alpha g_{10}\nu_0^3k^{4-d-y}+g_{20}\nu_0^3\right)\left|\frac{\epsilon_3(k)}{R(k)}\right|^2\right] \nonumber \\
&\times&\left[P_{bd}({\bm k})\frac{g_{10}\nu_0^3k^{4-d-y}+g_{20}\nu_0^3}{|\epsilon_1(k)|^2}+
Q_{bd}({\bm k})\left(\alpha g_{10}\nu_0^3k^{4-d-y}+g_{20}\nu_0^3\right)\left|\frac{\epsilon_3(k)}{R(k)}\right|^2\right];
\end{eqnarray}
\end{widetext}
hereinafter the Roman letters $i$ and $j$ are external (free) indices of the diagram, while the Roman letters $a,\ldots,d$
denote the vector indices of
the propagators with the implied summation over repeated indices. Two terms in the first line are vertices $V^1_{ijl}$ (see Fig.~\ref{fig:2}),
terms in  the second and the third line are propagators  {$\left\langle v_iv_j\right\rangle$}, see~\eqref{lines} and~\eqref{energies}. Since $d_\Gamma=0$ for this
diagram, we may put the external momenta ${\bm p}=0$.

The calculation of the tensor structure  {$J^1_{ij}$} gives
 {
\begin{eqnarray}
\label{J-D1}
J^1_{ij}=2(-\delta_{ij}k^2+k_i k_j) A(k)B(k),
\end{eqnarray}}
where $A(k)$ and $B(k)$ are the scalar parts of the propagators in the expression~\eqref{Expr-D1}, namely, 
\begin{eqnarray}
A(k)&=&\frac{g_{10}\nu_0^3k^{4-d-y}+g_{20}\nu_0^3}{|\epsilon_1(k)|^2}; \nonumber \\
B(k)&=&\left(\alpha g_{10}\nu_0^3k^{4-d-y}+g_{20}\nu_0^3\right)\left|\frac{\epsilon_3(k)}{R(k)}\right|^2.
\label{AB}
\end{eqnarray}

The integration over the frequency $\omega$ of the expression~\eqref{J-D1} gives 
\begin{eqnarray}
\int\frac{\dRM\omega}{2\pi}A(k)B(k)=\frac{1}{2k^6\nu_0^3u_0(u_0+1)},
\end{eqnarray}
therefore, the expression~\eqref{Expr-D1} takes the form
\begin{eqnarray}
\label{D1-tmp}
D_1&=&\int\frac{\dRM^d{\bm k}}{(2\pi)^d}\frac{\nu_0^3}{u_0(u_0+1)}\frac{1}{k^4} {\left(\delta_{ij}-\frac{k_i k_j}{k^2}\right)} \nonumber \\
&\times& \left(g_{10}k^{4-d-y}+g_{20}\right)\left(\alpha g_{10}k^{4-d-y}+g_{20}\right).
\end{eqnarray}
In order to integrate over the vector ${\bm k}$ we need to average the expression~\eqref{D1-tmp} over the angle variables:
\begin{equation}
\label{Int-Aver-Angles}
\int \dRM^d {\bm k} \, f({\bm k})=
S_{d} \int_m^\infty \dRM k\, k^{d-1}\,
\left\langle f({\bm k})\right\rangle,
\end{equation}
where $\langle\cdots\rangle$ is the averaging over the unit sphere in the
$d$-dimensional space, {$S_{d}=2\pi^{d/2}/\Gamma(d/2)$} is its surface area.
To perform an averaging of a given function of $\mk$ over
the angle variables we use the relations
\begin{align}
\left\langle \frac{k_i k_j}{k^2} \right\rangle
&= \frac{\delta_{ij}}{d}; \label{k-Angles0}\\
\left\langle \frac{k_i k_jk_lk_m}{k^4} \right\rangle
&= \frac{\delta_{ij}\delta_{lm}+\delta_{il}\delta_{jm}+\delta_{im}\delta_{jl}}{d(d+2)}.
\label{k-Angles}
\end{align}
In particular, Eq.~\eqref{k-Angles0} means that
\begin{eqnarray}
\int { \dRM^d {\bm k}} \frac{k_{i}k_{s}}{k^{2}} f(k) =
\frac{\delta_{is}}{d}\, \int { \dRM^d {\bm k}}\, f(k).
\end{eqnarray}
For $D_1$ this yields
\begin{eqnarray}
\label{D1-tmp2}
D_1&=&\frac{\nu_0^3}{u_0(u_0+1)}\frac{d-1}{d} {\delta_{ij}}C_d\int \dRM^d k\frac{ k^{d-1} }{k^4}\left[\alpha g^2_{10}k^{8-2d-2y}\right. \nonumber \\
&+& \left.(\alpha+1)g_{10}g_{20}k^{4-d-y} +g^2_{20}\right],
\end{eqnarray}
where $C_d=S_d/(2\pi)^d$. After the angular averaging has been performed, we are
left with simple integrals over the modulus $k$:
\begin{eqnarray}
\int_m^\infty \dRM^d k\, k^{d-1}\frac{k^{4-d-y}}{k^{4}} &=& \frac{m^{-y}}{y}; \nonumber \\
\int_m^\infty \dRM^d k\, k^{d-1}\frac{1}{k^{4}} &=& \frac{m^{-\varepsilon}}{\varepsilon},
\label{k-Int}
\end{eqnarray}
where $\varepsilon=4-d$. Applying these expressions {to} Eq.~\eqref{D1-tmp2}, one obtains
\pagebreak
\begin{widetext}
\begin{eqnarray}
\label{D1}
D_1=\frac{\nu_0^3}{u_0(u_0+1)}\frac{d-1}{d} {\delta_{ij}}C_d
\biggl[
   \frac{\alpha g_{10}^2}{2y-\eps}+\frac{(\alpha+1)g_{10}g_{20}}{y}+
   \frac{ g_{20}^2}{\eps}
 \biggl].
\end{eqnarray}

Taking into account the symmetry coefficient $1/2$ for this graph, Eq.~\eqref{eq:vsvs}
finally reads
\begin{eqnarray}
\Gamma_{v'v'} & =&  g_{10} \nu_0^3 \mu^y p^{4-d-y} \biggl\{
  P_{ij}({\bm p}) + \alpha Q_{ij}({\bm p})
  \biggl\} 
  + Z_6 g_{20} \nu_0^3 \delta_{ij} \nonumber \\
  &+& \frac{\nu_0^3}{u_0(u_0+1)}\frac{d-1}{2d}\delta_{ij}C_d
  \biggl[
   \alpha g_{10}^2\frac{m^{\eps-2y}}{2y-\eps}+(\alpha+1)g_{10}g_{20}\frac{m^{-y}}{y}+
   g_{20}^2\frac{m^{-\varepsilon}}{\eps}
 \biggl].
 \label{vprvpr-tmp}
\end{eqnarray}
\end{widetext}

\subsection{The diagrams with $d_\Gamma=1$}
\label{sec:Multiloop}

In this section we discuss now linearly divergent diagrams. We begin
with one of the diagrams, entering the expansion of the function $\left\langle \phi'v\right\rangle$ [see~\eqref{eq:phisv}], namely,
\begin{eqnarray}
D_2=\raisebox{-1ex}{ \includegraphics[width=2.5truecm]{PhisV1.eps}}.
\end{eqnarray}

In the frequency-momentum representation it is {given by}  \\
\begin{widetext}
\begin{align}
\label{Expr-D2}
D_2 &  =\int\frac{\dRM \omega}{2\pi}\int\frac{\dRM^d {\bm k}}{(2\pi)^d}V_b({\bm p+\bm k}) {V_j({\bm k})}\left\langle 
\phi\phi'\right\rangle(p+k)\left\langle  {v_b}\phi\right\rangle^*({\bm k}) \nonumber \\
 & = ic_0^2\nu_0^3\int\frac{\dRM \omega}{2\pi}\int\frac{ \dRM^d {\bm k}}{(2\pi)^d}
[k^2+({\bm p\cdot \bm k})] {k_j}\frac{\alpha g_{10}k^{4-d-y}+g_{20}}{\epsilon_2(k)\epsilon_3(k)\epsilon_3(p+k)\epsilon_2^*(k)},
\end{align}
\end{widetext}
where {both} $V_b({\bm p+\bm k})$ and $V_\beta({\bm k})$ are interaction vertices (see Fig.~\ref{fig:2}), 
$\left\langle \phi\phi'\right\rangle$ and $\left\langle  {v_b}\phi\right\rangle^*$ are two propagators, see~\eqref{lines}; ${\bm p}$ is an external
momenta, ${\bm k}$~-- internal one; $({\bm p\cdot \bm k})$ denotes the scalar product of vectors ${\bm p}$ and ${\bm k}$.

Since this diagram is linearly divergent, $d_\Gamma = 1$, only the terms proportional to ${\bm p}$ need to
be computed.
An integration of the scalar part in Eq.~\eqref{Expr-D2} over the frequency and an expansion of the result  
up to first order in~${\bm p}$ gives
\begin{align}
 \int\frac{\dRM \omega}{2\pi}
 &\frac{1}{\epsilon_2(k)\epsilon_3(k)\epsilon_3(p+k)\epsilon_2^*(k)} \nonumber \\ 
 &\cong\frac{1}{2u_0(u_0+v_0)^2\nu_0^3}\frac{1}{k^4}\left[\frac{1}{k^2}-\frac{2v_0({\bm p\cdot \bm k})}{k^4(u_0+v_0)}\right].
\end{align}
Substituting this expression into Eq.~\eqref{Expr-D2} and performing averaging over the angles [see Eq.~\eqref{k-Angles}] one obtains
\begin{eqnarray}
D_2&=& {p_j}\frac{1}{d}\frac{ic_0^2}{2u_0(u_0+v_0)^2}\left(1-\frac{2v_0}{u_0+v_0}\right)C_d \nonumber \\ 
&\times& \int k^{d-1} \dRM^d k\frac{1}{k^4}(\alpha g_{10}k^{4-d-y}+g_{20}).
\end{eqnarray}
Finally, use of Eq.~\eqref{k-Int} leads to the following result:
\begin{eqnarray}
\label{D2}
D_2&=&ic_0^2 {p_j}\frac{1}{d}\frac{u_0-v_0}{2u_0(u_0+v_0)^3}C_d \nonumber \\
&\times&\left(\alpha g_{10}\frac{m^{-y}}{y}+g_{20}\frac{m^{-\eps}}{\eps}\right).
\end{eqnarray}


The second diagram, entering the expansion of the function $\left\langle \phi'v\right\rangle$, is
\begin{eqnarray}
D_3=\raisebox{-1ex}{ \includegraphics[width=2.5truecm]{PhisV2.eps}}.
\end{eqnarray}

The analytical expression for the diagram reads
\begin{eqnarray}
\label{Expr-D3}
D_3&=&\int\frac{\dRM\omega}{2\pi}\int\frac{\dRM^d{\bm k}}{(2\pi)^d}V_a({\bm k+\bm p}) {V_{dcj}} \nonumber \\
&\times&\left\langle  {v_av_c}\right\rangle({\bm k})\left\langle \phi  {v_d'}\right\rangle({\bm p+\bm k}),
\end{eqnarray}
where  {$V_{dcj}$} and $V_a({\bm k+\bm p})$ are two vertices, $\left\langle  {v_av_c}\right\rangle$ and $\left\langle \phi  {v_d'}\right\rangle$ are two propagators.

The tensor structure  {$J^3_j$} for this diagram is
\begin{eqnarray}
 {J^3_j} &=&(-1)^2(k+p)_a( {k_j}\delta_{cd}+p_c {\delta_{dj}})(k+p)_d \nonumber \\
&\times&[P_{ac}({\bm k})A(k)+Q_{ac}({\bm k})B(k)],
\end{eqnarray}
where $A(k)$ and $B(k)$ are scalar coefficients from Eq.~\eqref{AB}. After summation over vector indices up to the
first order in ${\bm p}$ one obtains
\begin{eqnarray}
\label{D3-tmp}
 {J^3_j \cong k_j}[k^2+3({\bm p\cdot \bm k})]B(k).
\end{eqnarray}
Since we {have} put $c_0=0$ in all denominators, an integration over frequency and an
expansion of obtained expression up to first order in ${\bm p}$ yields
\begin{eqnarray}
\label{D3-tmp2}
\int\frac{\dRM \omega}{2\pi}
\frac{B(k)}{R(p+k)} = \int\frac{\dRM \omega}{2\pi} \frac{1}{|\epsilon_2(k)|^2R(p+k)} \nonumber \\
=\frac{1}{4\nu_0^3u_0^2(u_0+v_0)k^6}\left[1-\frac{u_0+3v_0}{u_0+v_0}\frac{({\bm p\cdot \bm k})}{k^2}\right].
\end{eqnarray}
Combining Eqs.~\eqref{D3-tmp} and~\eqref{D3-tmp2}, averaging the obtained result over angle variables, and
 applying Eq.~\eqref{k-Int} one 
 obtains
\begin{eqnarray}
\label{D3}
D_3&=&ic_0^2  {p_j}\frac{1}{d}\frac{1}{2u_0(u_0+v_0)^2}C_d \nonumber \\
&\times&\left(\alpha g_{10}\frac{m^{-y}}{y}+g_{20}\frac{m^{-\eps}}{\eps}\right).
\end{eqnarray}


The last diagram, entering the expansion of the function  {$\left\langle \phi'v_j\right\rangle$}, is
\begin{eqnarray}
D_4=\raisebox{-1ex}{ \includegraphics[width=2.5truecm]{PhisV3.eps}}.
\end{eqnarray}

The analytical expression for it is 
\begin{eqnarray}
\label{Expr-D4}
D_4&=&\int\frac{\dRM\omega}{2\pi}\int\frac{\dRM^d{\bm k}}{(2\pi)^d}V_a({\bm k}) {V_{dcj}} \nonumber \\
&\times&\left\langle  {v_d}\phi\right\rangle({\bm k})\left\langle  {v_a v_c'}\right\rangle({\bm p+\bm k}),
\end{eqnarray}
where  {$V_{dcj}$} and $V_a({\bm k})$ are two vertices, $\left\langle  {v_d}\phi\right\rangle$ and $\left\langle  {v_a v_c'}\right\rangle$ are two propagators.

The tensor structure  {$J^4_j$} for this diagram is
\begin{align}
 {J^4_j}& =k_ak_c( {k_j}\delta_{cd}+p_c {\delta_{dj}}) \nonumber \\
 \times&[P_{ad}({\bm p+\bm k})C(p+k)+Q_{ad}({\bm p+\bm k})D(p+k)],
\end{align}
where $C(p+k)$ and $D(p+k)$ are the scalar coefficients of the propagator $\left\langle  {v_a v_c'}\right\rangle$, namely,
\begin{eqnarray}
C(k)&=&\frac{1}{\epsilon_1(k)}; \quad 
D(k)=\frac{\epsilon_3(k)}{R(k)}.
\label{CD}
\end{eqnarray}
After the summation over vector indices up to the first order {in} ${\bm p}$ one obtains
\begin{eqnarray}
\label{D4-tmp}
 {J^4_j} &\cong&  {k_j}[k^2+({\bm p\cdot \bm k})]D(p+k).
\end{eqnarray}
Integration over the frequency of the scalar part of the expression~\eqref{Expr-D4} gives 
\begin{align}
\label{D4-tmp2}
\int\frac{\dRM\omega}{2\pi}
 &\frac{D(p+k)\epsilon_3(k)}{|R(k)|^2} = \int\frac{\dRM\omega}{2\pi}\frac{1}{\epsilon_2(p+k)|\epsilon_2(k)|^2\epsilon_3^*(k)} \nonumber \\
&=\frac{1}{2\nu_0^3u_0^2(u_0+v_0)k^4} \nonumber \\
&\times \frac{u({\bm p+\bm k})^2+k^2(2u_0+v_0)}{[k^2+({\bm p+\bm k})^2][v_0k^2+u_0({\bm p+\bm k})^2]}.
\end{align}
In the same way as it has been done previously we obtain the following result:
\begin{equation}
\label{D4}
D_4=-ic_0^2  {p_j}\frac{1}{d}\frac{1}{(u_0+v_0)^3}C_d 
\left(\alpha g_{10}\frac{m^{-y}}{y}+g_{20}\frac{m^{-\eps}}{\eps}\right).
\end{equation}

From Eqs.~\eqref{D2},~\eqref{D3} and~\eqref{D4} it follows that 
\begin{eqnarray}
D_2+D_3+D_4=0
\end{eqnarray}
in the first order {in} $g_1$ and $g_2$. From this fact we immediately {conclude}~[see~\eqref{eq:phisv}] that 
\begin{eqnarray}
\label{Z5}
Z_5=1.
\end{eqnarray}
Unlike the functions  {$\left\langle v_i'v_jv_k\right\rangle$} and, for example,  {$\left\langle \phi'v_iv_j\right\rangle$} (see~Sec.~\ref{sec:Canon}), the finiteness of the function 
 {$\left\langle \phi'v_j\right\rangle$} is not {because of} an internal symmetry of the system, but it is the result of direct
calculations, i.e., the result of cancellation of the non-trivial contributions of three diagrams. Therefore, it is unclear 
whether this result is exact or it is broken in higher orders of the perturbation theory.


The last diagram with $d_\Gamma=1$ is the diagram, entering the expansion of the function  {$\left\langle v_i'\phi\right\rangle$} 
 [see~\eqref{eq:vsphi}], namely,
\begin{eqnarray}
D_5=\raisebox{-1ex}{ \includegraphics[width=2.5truecm]{PhiVs.eps}}.
\end{eqnarray}

The analytical expression for it is 
\begin{equation}
\label{Expr-D5}
D_5 = \int\frac{\dRM \omega}{2\pi}\int\frac{\dRM^d{\bm k}}{(2\pi)^d} {V_{iab}}V_c({\bm p}) 
\left\langle  {v_av_c}\right\rangle({\bm k})\left\langle  {v_b}\phi'\right\rangle({\bm p+\bm k}),
\end{equation}
where  {$V_{iab}$} and $V_c({\bm p})$ are two vertices, $\left\langle  {v_av_c}\right\rangle$ and $\left\langle  {v_b}\phi'\right\rangle$ are two propagators.

The tensor structure  {$J^5_i$} for this diagram is
\begin{eqnarray}
 {J^5_i}&=&(-i)^3[k_b {\delta_{ia}}-(p+k)_a {\delta_{ib}}](-p_c)(k+p)_b \nonumber \\
&\times&[P_{ac}({\bm k})A(k)+Q_{ac}({\bm k})B(k)],
\end{eqnarray}
where $A(k)$ and $B(k)$ are the scalar coefficients~\eqref{AB}. After the summation over the vector indices, up to 
the first order in ${\bm p}$ one obtains
\begin{eqnarray}
\label{D5-tmp}
 {J^5_i} \cong[ {p_i}k^2-({\bm p\cdot \bm k}) {k_i}]A(k).
\end{eqnarray}
Since we are {interested} only in the terms proportional to ${\bm p}$ and the expression~\eqref{D3-tmp} does not contain zero
order term ${\bm p}^0$, we may put ${\bm p}=0$ in all denominators; hence, integration over the frequency gives 
\begin{eqnarray}
\int\frac{\dRM \omega}{2\pi}
\frac{A(k)}{R(k)} = \frac{1}{2\nu_0^3(u_0+1)(v_0+1)k^6}.
\end{eqnarray}
Finally, using the formulas~\eqref{k-Angles} and~\eqref{k-Int} one obtains the following result:
\begin{eqnarray}
\label{D5}
D_5&=&-i  {p_i}\frac{d-1}{d}\frac{1}{2(u_0+1)(v_0+1)}C_d \nonumber \\
&\times&\left(g_{10}\frac{m^{-y}}{y}+g_{20}\frac{m^{-\eps}}{\eps}\right).
\end{eqnarray}

\subsection{The diagrams with $d_\Gamma=2$}
\label{sec:TruePropagator}

Start with the diagram, entering the expansion for function $\left\langle \phi\phi'\right\rangle$ 
[see~\eqref{eq:phiphis}], namely,
\begin{eqnarray}
D_6=\raisebox{-1ex}{ \includegraphics[width=2.5truecm]{PhiPhiS.eps}}.
\end{eqnarray}

The analytical expression for it is 
\begin{eqnarray}
\label{Expr-D6}
D_6&=&\int\frac{\dRM\omega}{2\pi}\int\frac{\dRM^d {\bm k}}{(2\pi)^d}V_{a}({\bm p+\bm k})V_c({\bm p}) 
\nonumber \\ &\times&\left\langle \phi\phi'\right\rangle(p+k)\left\langle  {v_av_c}\right\rangle({\bm k}),
\end{eqnarray}
where $V_{a}({\bm k})$ and $V_c({\bm k})$ are two vertices, $\left\langle \phi\phi'\right\rangle$ and $\left\langle  {v_av_c}\right\rangle$ are two propagators.

We are {interested} in the term,  proportional to $p^2$, therefore, the tensor structure $J^6$ for this diagram is
\begin{eqnarray}
J^6 &=&\left[p^2k^2-({\bm p\cdot \bm k})^2\right]A(k)  \nonumber \\
&+& \left[({\bm p\cdot \bm k})k^2+({\bm  p\cdot \bm k})^2\right]B(k),
\end{eqnarray}
where $A(k)$ and $B(k)$ are the scalar coefficients~\eqref{AB}. After integration over the frequency with 
function $1/\epsilon_3(p+k)$, which came from the propagator 
$\left\langle \phi\phi'\right\rangle$, using the formulas~\eqref{k-Angles} and~\eqref{k-Int} one obtains the following expression:
\begin{eqnarray}
\label{D6}
D_6&=&
-\frac{\nu_0 }{2d}p^2C_d\left[
\frac{d-1}{1+v_0} 
\biggl(
 g_{10}\frac{m^{-y}}{y} + g_{20}\frac{m^{-\eps}}{\eps}
 \biggl)\right. \nonumber \\
  &+& \left.\frac{(u_0-v_0)}{u_0(u_0+v_0)^2} \biggl( 
 \alpha g_{10}\frac{m^{-y}}{y} + g_{20}\frac{m^{-\eps}}{\eps}
 \biggl) \right].
\end{eqnarray}


The last diagram $D_7$, which {enters the expansion} of the function  {$\left\langle v_i'v_j\right\rangle$}, namely, 
\begin{eqnarray}
D_7=\raisebox{-1ex}{ \includegraphics[width=2.5truecm]{VVs.eps}},
\end{eqnarray}
is more complicated~-- both fields $ {v_i}$ and $ {v_i'}$ are vector fields, therefore, there are two
possible structures  {$p^2\delta_{ij}$} and  {$p_i p_j$}, which are both quadratic in external momentum; see~\eqref{eq:phiphis}.
The analytical expression for it is
\begin{equation}
\label{Expr-D7}
D_7 =\int\frac{\dRM \omega}{2\pi}\int\frac{ \dRM^d {\bm k}}{(2\pi)^d} {V_{iab}V_{jcd}} 
\left\langle  {v_av_c}\right\rangle({\bm k})\left\langle  {v_bv_d'}\right\rangle({\bm p-\bm k}),
\end{equation}
where  {$V_{iab}$} and  {$V_{jcd}$} are two vertices $V^1_{ijl}$ (see Fig.~\ref{fig:2}), $\left\langle  {v_av_c}\right\rangle$ and 
$\left\langle  {v_bv_d'}\right\rangle$ are two propagators.

Divide the expression~\eqref{Expr-D7} into four parts and calculate them separately:
\begin{eqnarray}
D_7&=&\int\frac{ \dRM\omega}{2\pi}\int\frac{ \dRM^d {\bm k}}{(2\pi)^d}[k_b {\delta_{ia}}+(p-k)_a {\delta_{ib}}] \nonumber \\
&\times&(-p_c {\delta_{jd}}+ {k_j}\delta_{cd}) \left(I_1+I_2+I_3+I_4\right),
\end{eqnarray}
where
\begin{eqnarray}
I_1&=&P_{ac}({\bm k})P_{bd}({\bm p-\bm k}) A(k)C(p-k); \nonumber \\
I_2&=&P_{ac}({\bm  k})Q_{bd}({\bm  p-\bm k}) A(k)D(p-k); \nonumber \\
I_3&=&Q_{ac}({\bm  k})P_{bd}({\bm  p-\bm k}) B(k)C(p-k); \nonumber \\
I_4&=&Q_{ac}({\bm  k})Q_{bd}({\bm  p-\bm k}) B(k)D(p-k);
\end{eqnarray}
see~\eqref{lines},~\eqref{energies},~\eqref{AB}, and~\eqref{CD}.

After integration over the internal frequency $\omega$, expanding the obtained result in external momentum ${\bm p}$ up to
second order, averaging it over the angle variables~[see~\eqref{k-Angles}] and integrating over the modulus $k$~[see~\eqref{k-Int}] 
one obtains 
\begin{widetext}
\begin{eqnarray}
\label{I1}
\hat{I}_1&=&\frac{1}{4}\nu_0C_d\left[p^2 {P_{ij}({\bm p})}\frac{1-d}{d+2}- {p_i p_j}\frac{2(d-1)^2}{d(d+2)}\right]
\biggl(  g_{10}\frac{m^{-y}}{y} + g_{20}\frac{m^{-\eps}}{\eps} \biggl); \\
\hat{I}_2&=&\frac{1}{2(u_0+1)}\nu_0C_d\left[p^2 {P_{ij}({\bm p})}\left(\frac{4+8u_0}{1+u_0}\frac{1}{d(d+2)}-\frac{2}{d}\right)
+ {p_i p_j}\frac{d-1}{d}\left(1-\frac{4+8u_0}{1+u_0}\frac{1}{d+2}\right)\right]
\nonumber \\
&\times& \biggl(  g_{10}\frac{m^{-y}}{y} + g_{20}\frac{m^{-\eps}}{\eps} \biggl);\\
\label{I2}
\hat{I}_3&=&\frac{1-u_0}{2u_0(1+u_0)^2}\nu_0C_d\frac{1}{d} {P_{ij}({\bm p})}
\biggl(\alpha g_{10}\frac{m^{-y}}{y} + g_{20}\frac{m^{-\eps}}{\eps} \biggl); \\
\label{I3}
\hat{I}_4&=&0,
\label{I4}
\end{eqnarray}
where
\begin{eqnarray}
\hat{I}_i&=&\int\frac{\dRM\omega}{2\pi}\int\frac{\dRM^d{\bm k}}{(2\pi)^d}[k_b {\delta_{ia}}+(p-k)_a {\delta_{ib}}] 
(-p_c {\delta_{jd}}+ {k_j}\delta_{cd})  I_i.
\end{eqnarray}

Combination of the expressions~\eqref{I1}~-- \eqref{I4} together leads to the following result for the diagram $D_7$:
\begin{eqnarray}
\label{D7}
D_7&=&p^2 {P_{ij}({\bm p})}I_\perp +  {p_i p_j} I_{\parallel},
\end{eqnarray}
where
\pagebreak
\begin{align}
 I_{\perp}  = -\frac{\nu_0 C_d}{2d(1+u_0)^2} & \left[
   \frac{u_0^2d(d-1)+u_0(2d^2+2d-8)+d(d+3)}{2(d+2)}
 \biggl( g_{10}\frac{m^{-y}}{y} + g_{20}\frac{m^{-\eps}}{\eps} \biggl) \right.
  \nonumber\\
& \left.+  \frac{(u_0-1)}{u_0}
\biggl(\alpha g_{10}\frac{m^{-y}}{y} + g_{20}\frac{m^{-\eps}}{\eps} \biggl)\right]; \\
 I_{\parallel} =-\frac{\nu_0C_d}{2d(1+u_0)^2} & (d-1)
 \frac{u_0^2(d-1) +u_0(d+4) + 1}{(d+2)}
 \biggl(g_{10}\frac{m^{-y}}{y} + g_{20}\frac{m^{-\eps}}{\eps} \biggl).
\label{D7-Ipar}
\end{align}
\end{widetext}

The expressions~\eqref{D1},~\eqref{D5},~\eqref{D6}, and~\eqref{D7} are final answers
for divergent parts of all the Green functions, which are needed to renormalize the model. 
To find renormalization constants it is necessary to put them into the expressions~\eqref{eq:vsv}~-- \eqref{eq:vsvs} and to require their UV 
finiteness (when they are expressed in new renormalized variables), i.e., finiteness at $y\to0$ and $\varepsilon\to0$.

\subsection{Renormalization constants $Z_1$~-- $Z_6$}

From the expressions~\eqref{eq:vsv} and~\eqref{D7}~-- \eqref{D7-Ipar} it follows that the
renormalization constant $Z_1$ is connected {to} the
expression $ I_{\perp}$, while the
renormalization constant $Z_2$ is connected {to} the expression $ I_{\parallel}$. Moreover, one should not forget all
 the factors like $u_0$, $v_0$ or $g_{1/2,0}$, which
are presented in the terms of the action functional and are not necessary presented in the results of calculations of diagrams; see, for 
example, expression~\eqref{vprvpr-tmp}~-- not all the terms in the expression~\eqref{D1} are proportional to the coupling constant $g_{20}$.
 {In the one-loop approximation we may always replace the bare couplings $g_{1/2,0}$ by their renormalized counterparts $g_{1/2}$: since we have
 already singled out poles in $y$ and $\eps$, taking into account corrections $Z_{g_{1/2}}$ would be an excess of accuracy; see, e.g.,~\eqref{D2}. 
The multipliers like $(m/\mu)^y$, which are connected with $Z_{g_{1/2}}$, in the MS scheme are equal to $1$.
This observation also takes place for all other parameters, namely $u$, $v$, $\nu$, and $c$.}
Passing to new variables according {to the convention} $g_{1,2}\to g_{1,2}C_d$ one finally obtains
\begin{align}
  Z_1 & = 1 -  
 \frac{u^2d(d-1)+u(2d^2+2d-8)+d(d+3)}{4d(d+2)(1+u)^2}
  \nonumber \\
 &\times \biggl( 
 \frac{g_{1} }{y} + \frac{g_{2} }{\eps}
 \biggl)
 -  \frac{u-1}{2du(1+u)^2}
 \biggl( \alpha 
 \frac{g_{1}}{y} + \frac{g_{2} }{\eps}
 \biggl);  \nonumber \\
  Z_2 & = 1 - (d-1)
 \frac{u^2(d-1) +u(d+4) + 1}{2d(d+2)u(1+u)^2}\biggl(
 \frac{g_{1}}{y} + \frac{g_{2} }{\eps}
 \biggl); \nonumber \\
  Z_3 & = 1 -\frac{1}{2dv}\biggl[
\frac{d-1}{v+1} 
\biggl(
 \frac{g_{1}}{y} + \frac{g_{2}}{\eps}
 \biggl)
 \nonumber\\
 & +
 \frac{(u-v)}{u(u+v)^2} \biggl( \alpha 
 \frac{g_{1}}{y} + \frac{g_{2}}{\eps}
 \biggl) 
 \biggl];\nonumber \\
  Z_4 & = 1 + \frac{d-1}{2 d(1+u)(1+v)}
 \biggl(
 \frac{g_{1}}{y} + \frac{g_{2} }{\eps}
 \biggl); \nonumber \\
  Z_6 & = 1 -\frac{d-1}{2d u(1+u)}
 \biggl[
  \alpha  \frac{g_{1}^2}{g_2(2y-\eps)}+(\alpha+1)\frac{g_{1}}{y}+
   \frac{ g_{2}}{\eps}
 \biggl].
\end{align}
As it was mentioned {in~\eqref{Z5}}, the renormalization constant $Z_5$ is trivial.

To find renormalization constants $Z$ for the fields $\phi$ and $\phi'$ and physical parameters of the system on should use
the relations~\eqref{eq:RGconst2} and the binomial relation $(1+x)^{-n}=1-nx+\mathcal{O}(x^2)$, which is necessary to calculate $Z_i^{-n}$.

\section{Calculation of the diagrams for advection-diffusion stochastic equation}
\label{app2}
In this section the detailed calculations of the diagrams, defining the renormalization constants $Z_\kappa$ (see Sec.~\ref{sec:AmonDim}) and $Z_n$ (see Sec.~\ref{sec:Ops}), are presented.

\subsection{The diagram for response function $\left\langle \theta'\theta\right\rangle$}
\label{app2a}

The constant $Z_{\kappa}$ is to be found from the
requirement of UV finiteness of the 1-irreducible Green function
$\langle\theta'\theta\rangle$.
Like for the original Navier-Stokes model, 
the divergent part of the considered Feynman diagram is independent on $c_{0} \sim c$ and, therefore, can be calculated
directly at $c=0$. 

An analytical expression for the diagram $D_8$, 
\begin{eqnarray}
D_8=\raisebox{-1ex}{ \includegraphics[width=2.5truecm]{G2_1.eps}},
\end{eqnarray}
which enters the expression~\eqref{Dyson}, is
\begin{eqnarray}
\label{Expr-D8}
D_8&=&\int\frac{\dRM\omega}{2\pi}
\int\frac{ \dRM^d {\bm k}}{(2\pi)^{d}} V_{a}({\bm p})V_{c}({\bm p+\bm k})
\frac{1}
{-i\omega+w_0\nu_0 ({\bm p}+{\bm k})^{2}} \nonumber \\
&\times&\left[P_{ac}({\bm k})  A(k)+Q_{ac}({\bm k})  B(k)\right].
\end{eqnarray}
Here $V_{a}({\bm p})=ip_a$ and $V_c({\bm p+\bm k})=i(p+k)_c$ are two vertices of the type~\eqref{vertex1}, {$w_0=\nu_0/\kappa_0$}; 
scalar coefficients $A(k)$ and $B(k)$ of the propagator  {$\left\langle v_av_c\right\rangle$} are defined in~\eqref{AB}.

Since in the leading-order approximation the renormalization constant $Z_\kappa$ in the
bare term of~(\ref{Dyson}) is taken only in the first order in coupling constants $g_1$ and $g_2$, i.e.,
$\kappa_{0} = \kappa Z_{\kappa} \simeq \kappa (1 + z^{(1)}_y g_1/y+ z^{(1)}_\eps g_2/\eps)$,
during the actual calculation all other renormalization constants
in the diagram $D_8$, entering, for example, the functions $A(k)$ and $B(k)$, should be replaced with unities.

From the integration over the frequency we get

\begin{align}
\int\frac{\dRM\omega}{2\pi} \frac{A(k)}
{-i\omega+w\nu ({\bm p}+{\bm k})^{2}} &= \frac{1}{2\nu^2k^2[k^2+w({\bm p+\bm k})^2]}; \nonumber \\
\int\frac{\dRM\omega}{2\pi} \frac{B(k)}
{-i\omega+w\nu ({\bm p}+{\bm k})^{2}} &= \frac{1}{2\nu^2uk^2[uk^2+w({\bm p+\bm k})^2]}.
\label{Tmp}
\end{align}

The expression~\eqref{Expr-D8} can be separated into two parts: 
\begin{align}
\check{I}_1&=-\int\frac{\dRM^d{\bm k}}{(2\pi)^{d}}p_a(p+k)_c  \frac{P_{ac}({\bm k})}{2\nu^2k^2[k^2+w({\bm p+\bm k})^2]}; \nonumber \\
\check{I}_2&=-\int\frac{\dRM^d{\bm k}}{(2\pi)^{d}}p_a(p+k)_c  \frac{Q_{ac}({\bm k})}{2\nu^2uk^2[uk^2+w({\bm p+\bm k})^2]}. \nonumber\\
\end{align}

We are interested in the term  proportional to $p^2$. Therefore, in computation
of $\check{I}_1$ one may
immediately set ${\bm p}=0$ and, using the expressions~\eqref{k-Angles}, \eqref{k-Int}, and the notation $g_iC_d\to g_i$, get
\begin{align}
\check{I}_1&=-p^2\frac{\nu}{2(w+1)}\frac{d-1}{d}\biggl(
 g_{1}\frac{m^{-y}}{y} + g_{2}\frac{m^{-\eps}}{\eps}
 \biggl).
\end{align}
Using Taylor expansion up to the linear term in $p$ for the second part of~\eqref{Tmp} we can rewrite $\check{I}_2$ in the form

\begin{align}
\check{I}_2&=-p^2\frac{\nu}{2u(u+w)}\frac{1}{d}\frac{u-w}{u+w}\biggl(
\alpha g_{1}\frac{m^{-y}}{y} + g_{2}\frac{m^{-\eps}}{\eps}
 \biggl).
\end{align}
Finally, collecting $\check{I}_1$ and $\check{I}_2$ yields
\begin{widetext}
\vspace{-1.5em}
\begin{align}
\label{D8}
D_8=
-p^2\frac{\nu}{2}\frac{1}{d} 
\left[\biggl(
 \frac{d-1}{w+1}+ \frac{\alpha}{u(u+w)}- 2\frac{\alpha w}{u(u+w)^2}
 \biggl) \left(\frac{\mu}{m}\right)^y \frac{ g_1 }{y}
-\biggl(
 \frac{d-1}{w+1}+ \frac{1}{u(u+w)}- 2\frac{w}{u(u+w)^2}
 \biggl) \left(\frac{\mu}{m}\right)^\varepsilon\frac{ g_2 }{\varepsilon}\right].
\end{align}
\vspace{-1.5em}

Therefore, the renormalization constant $Z_\kappa$ [see~\eqref{Dyson}] should be chosen as
\begin{align}
Z_{\kappa} = 1 - \frac{1}{2dw}
\left[ \frac{d-1}{w+1} + \frac{\alpha(u-w)}{u(u+w)^{2}}
\right]\frac{g_1}{y} 
- \frac{1}{2dw}
\left[ \frac{d-1}{w+1} + \frac{u-w}{u(u+w)^{2}}
\right]\frac{g_2}{\varepsilon}.
\label{Zk}
\end{align}
\end{widetext}

\subsection{The diagram for composite operator $\theta^n(x)$}
\label{app2b}

The divergence of the graph $D_9$,
\begin{align}
D_9=  \vcenter{\hbox
{\includegraphics [width=.09\textwidth,clip]{comp_density.eps}}},
\end{align}
entering into the expansion~\eqref{Diag-General}, is logarithmic, hence, one might set all the external
frequencies and momenta equal to zero. Therefore, the analytical expression of the diagram is given by
\begin{eqnarray}
\label{D9-Expr}
D_9&=&\int\frac{\dRM\omega}{2\pi}
\int\frac{\dRM^d{\bm k}}{(2\pi)^{d}} V_{a}({\bm k})V_{c}({-\bm k})
\frac{1}{\omega^{2}+w^{2}\nu^2 k^{4}} \nonumber \\
&\times&\left[P_{ac}({\bm k}) A(k)+Q_{ac}({\bm k})  B(k)\right],
\end{eqnarray}
where $V_{a}({\bm k})$ and $V_c(-{\bm k})$ are two vertices~\eqref{vertex1}; 
scalar coefficients $A(k)$ and $B(k)$ of the propagator  {$\left\langle v_av_c\right\rangle$} are defined in~\eqref{AB} with the replacement of
original bare parameters to their renormalized counterparts. As $V_{a}({\bm k})P_{ac}({\bm k})=0$, only the second term in~\eqref{D9-Expr} gives a non-vanishing contribution. 

Integration over the frequency gives
\begin{align}
\int\frac{\dRM \omega}{2\pi} \frac{B(k)}{\omega^{2}+w^{2}\nu^2 k^{4}} &= \frac{1}{2\nu^3}\frac{1}{uw(u+w)}\frac{1}{k^6}.
\label{Tmp2}
\end{align}
Contracting tensor indices, using~\eqref{k-Int}, and collecting all the factors the expression~\eqref{Gamma2} can be
rewritten as follows
\begin{align}
\Gamma_{n}(x;\theta)&= \theta^{n}(x)  \biggl\{ 1+
\frac{n(n-1)}{4wu(u+w)}
\biggl[\alpha g_1\left(\frac{\mu}{m}\right)^y\frac{1}{y}&\nonumber\\
&+g_2\left(\frac{\mu}{m}\right)^\varepsilon\frac{1}{\varepsilon}\biggl] \biggl\},
\label{Gamma22}
\end{align}
where the substitution $g_i\to g_iC_d$ is implied.

The renormalization constants $Z_{n}$ are found from the requirement that the
renormalized analog $\Gamma_{n}^{R}=Z_{n}^{-1}\Gamma_{n}$ of the function
(\ref{Gamma1}) be UV finite in terms of renormalized parameters.
In contrast to the expressions~\eqref{eq:vsv}~-- \eqref{eq:vsvs}, in this case the renormalization constants $Z_n$
do not pertain to some model parameters, but to the Green functions themselves.
Hence, using the loop expansion~\eqref{Gamma2}
one does not find the renormalization constants $Z_n$, but {an inversed one} $Z_n^{-1}$.
Taking into account a minus sign in the exponent,
from~\eqref{Gamma22} it follows that in the MS scheme the renormalization constants take a form
\begin{equation}
Z_{n} = 1 + \frac{n(n-1)}{4wu(u+w)}\left(\frac{\alpha g_1}{y}+\frac{g_2}{\varepsilon} \right).
\label{Zn}
\end{equation}


\end{document}